%% file: paper.tex
\documentclass[default,iicol]{sn-jnl}

\usepackage{graphicx}%
\usepackage{multirow}%
\usepackage{amsmath,amssymb,amsfonts}%
\usepackage{amsthm}%
\usepackage{mathrsfs}%
\usepackage[title]{appendix}%
\usepackage{xcolor}%
\usepackage{textcomp}%
\usepackage{manyfoot}%
\usepackage{booktabs}%
\usepackage{algorithm}%
\usepackage{algorithmicx}%
\usepackage{algpseudocode}%
\usepackage{listings}%

\theoremstyle{thmstyleone}%
\theoremstyle{thmstyletwo}%

\theoremstyle{thmstylethree}%

\usepackage[utf8]{inputenc}
\usepackage[scientific-notation=true]{siunitx}
\usepackage{import}
\usepackage{tikz}

\usepackage{pgfplots}
\pgfplotsset{compat=1.12}
\usepgfplotslibrary{fillbetween}
\usepgfplotslibrary{groupplots}
\usepgfplotslibrary{colormaps}
\usetikzlibrary{positioning}

\include{figures/images/cet_linear_worb_100_25_c53}

\usepackage{nicefrac}
\usepackage{xcolor}
\definecolor{dia3x1}{HTML}{ffa600}
\definecolor{dia3x2}{HTML}{bc5090}
\definecolor{dia3x3}{HTML}{003f5c}

\definecolor{dia4x1}{HTML}{003f5c}
\definecolor{dia4x2}{HTML}{7a5195}
\definecolor{dia4x3}{HTML}{ef5675}
\definecolor{dia4x4}{HTML}{ffa600}

\definecolor{dia5x1}{HTML}{003f5c}
\definecolor{dia5x2}{HTML}{58508d}
\definecolor{dia5x3}{HTML}{bc5090}
\definecolor{dia5x4}{HTML}{ff6361}
\definecolor{dia5x5}{HTML}{ffa600}

\definecolor{diav1x1}{HTML}{003f5c}
\definecolor{diav1x2}{HTML}{8699aa}
\definecolor{diav2x1}{HTML}{ffa600}
\definecolor{diav2x2}{HTML}{ffd291}

\definecolor{dia1-1}{HTML}{001f2e} 
\definecolor{dia1-2}{HTML}{003f5c}
\definecolor{dia1-3}{HTML}{0095d9} 

\definecolor{dia2-1}{HTML}{873365} 
\definecolor{dia2-2}{HTML}{bc5090}
\definecolor{dia2-3}{HTML}{ffd6ed}

\definecolor{dia3-1}{HTML}{cc8400} 
\definecolor{dia3-2}{HTML}{ffa600}
\definecolor{dia3-3}{HTML}{fbddbe}

\definecolor{dia6-1-1}{HTML}{001F2E}
\definecolor{dia6-1-2}{HTML}{274758}
\definecolor{dia6-1-3}{HTML}{4E6F82}
\definecolor{dia6-1-4}{HTML}{7497AB}
\definecolor{dia6-1-5}{HTML}{9BBFD5}
\definecolor{dia6-1-6}{HTML}{C2E7FF}

\definecolor{dia6-2-1}{HTML}{873365}
\definecolor{dia6-2-2}{HTML}{9B4E7C}
\definecolor{dia6-2-3}{HTML}{AF6992}
\definecolor{dia6-2-4}{HTML}{C385A9}
\definecolor{dia6-2-5}{HTML}{D7A0C0}
\definecolor{dia6-2-6}{HTML}{FFD6ED}

\definecolor{dia6-3-1}{HTML}{CC8400}
\definecolor{dia6-3-2}{HTML}{D59626}
\definecolor{dia6-3-3}{HTML}{DFA84C}
\definecolor{dia6-3-4}{HTML}{E8B972}
\definecolor{dia6-3-5}{HTML}{F2CB98}
\definecolor{dia6-3-6}{HTML}{FBDDBE}

\newcommand*{\pd}{\partial}
\newcommand*{\dd}{\mathrm{d}}
\newcommand*{\bs}[1]{\boldsymbol{#1}}
\newcommand*{\Trans}{\mathrm{T}}

\newcommand*{\plotquantiles}[4][]{
    \addplot[name path=A,color=#3!70, forget plot, line width=0.75pt, #1, densely dotted] table [x=#4, y=quantile_0.99, col sep=comma]
        {#2};
    \addplot[name path=B,color=#3!70, forget plot, line width=0.75pt, #1, densely dotted] table [x=#4, y=quantile_0.01, col sep=comma]
        {#2};

    \addplot[mark=none, line width=1pt, color=#3, #1]
    table [x=#4, y=average, col sep=comma]
        {#2};
}

\input{writing_utils}
\begin{document}

\title{Adaptive integration of history variables in constrained mixture models for organ-scale growth and remodeling}


\author*[1]{\fnm{Amadeus M.} \sur{Gebauer}}\email{amadeus.gebauer@tum.de}

\author[2]{\fnm{Martin R.} \sur{Pfaller}}\email{pfaller@stanford.edu}

\author[3]{\fnm{Jason M.} \sur{Szafron}}\email{jszafron@andrew.cmu.edu}

\author[1,4]{\fnm{Wolfgang A.} \sur{Wall}}\email{wolfgang.a.wall@tum.de}

\affil[1]{\orgdiv{Institute for Computational Mechanics}, \orgname{Technical University of Munich}, \orgaddress{\street{Boltzmannstr. 15}, \city{Garching}, \postcode{85748}, \country{Germany}}}

\affil[2]{\orgdiv{Pediatric Cardiology, Stanford Maternal \& Child Health Research Institute, and Institute for Computational and Mathematical Engineering}, \orgname{Stanford University}, \orgaddress{\city{Stanford}, \country{USA}}}

\affil[3]{\orgdiv{Department of Biomedical Engineering}, \orgname{Carnegie Mellon University}, \orgaddress{\city{Pittsburgh}, \country{USA}}}

\affil[4]{\orgdiv{Munich Institute of Biomedical Engineering}, \orgname{Technical University of Munich}, \orgaddress{\street{Boltzmannstr. 11}, \city{Garching}, \postcode{85748}, \country{Germany}}}


\abstract{
    In the last decades, many computational models have been developed to predict
    soft tissue growth and remodeling (G\&R). The constrained
    mixture theory describes fundamental mechanobiological processes in soft tissue G\&R and    
    has been widely adopted in cardiovascular models of G\&R. However, even after two decades of work,
    large organ-scale models are rare, mainly due to
    high computational costs (model evaluation and memory consumption),
    especially in long-range simulations.
    We propose two strategies to adaptively integrate history variables in constrained mixture models
    to enable large organ-scale simulations of G\&R. Both strategies exploit that the influence of
    deposited tissue on the current mixture decreases over time through degradation. One strategy
    is independent of external loading, allowing the estimation of the computational resources ahead
    of the simulation. The other adapts the history snapshots based on the local mechanobiological environment so that
    the additional integration errors can be controlled and kept negligibly small, even in G\&R scenarios with severe perturbations.
    We analyze the adaptively integrated constrained mixture model on a tissue patch for a parameter study and
    show the performance under different G\&R scenarios. 
    To confirm that adaptive strategies enable large organ-scale
    examples, we show simulations of different hypertension conditions with a real-world example of a
    biventricular heart discretized with a finite element mesh.
    In our example, adaptive integrations sped up simulations by a factor of three and reduced
    memory requirements to one-sixth. The reduction of the computational costs gets even more pronounced
    for simulations over longer periods.
    Adaptive integration of the history variables allows studying more finely resolved models and
    longer G\&R periods while computational costs are drastically reduced and largely constant in time.
}




\maketitle

\input{sections/introduction}
\input{sections/modeling}
\input{sections/results_1D}
\input{sections/results_3D}
\input{sections/discussion}

\appendix

\input{sections/appendix}

\backmatter








\section*{Declarations}

\bmhead{Author contributions}
AMG was involved in conceptualization, methodology, software, validation visualization, and original draft writing.
MRP contributed to methodology, validation, visualization, and draft review and editing.
JMS was involved in validation, visualization, and reviewing and editing the draft.
WAW contributed to conceptualization, funding acquisition, resources, supervision, and draft review and editing.

\bmhead{Funding}

MRP was supported by the National Heart, Lung, and Blood Institute of the National Institutes of
Health under Award Number K99HL161313 and the Stanford Maternal and Child Health Institute. The
content is solely the responsibility of the authors and does not represent the official views of the
National Institutes of Health. WAW acknowledges the support by BREATHE, a Horizon 2020-ERC-2020-ADG
project, grant agreement No. 101021526-BREATHE.


\bmhead{Conflict of interest}
The authors declare that they have no competing interests.

\bmhead{Ethical approval}
not applicable





\bibliographystyle{sn-basic}
\bibliography{references}

\end{document}

%% file: figures/images/cet_linear_worb_100_25_c53.tex
\pgfplotsset{
  colormap={cetlinearworb}{
    rgb255=(254,255,255)
    rgb255=(254,253,249)
    rgb255=(252,251,240)
    rgb255=(251,249,235)
    rgb255=(250,247,226)
    rgb255=(249,245,221)
    rgb255=(247,243,212)
    rgb255=(245,241,204)
    rgb255=(244,239,199)
    rgb255=(242,237,190)
    rgb255=(242,235,187)
    rgb255=(242,232,182)
    rgb255=(242,229,177)
    rgb255=(242,226,173)
    rgb255=(242,223,168)
    rgb255=(242,221,165)
    rgb255=(242,218,160)
    rgb255=(242,216,157)
    rgb255=(241,213,152)
    rgb255=(242,209,148)
    rgb255=(242,207,146)
    rgb255=(242,204,142)
    rgb255=(243,201,140)
    rgb255=(243,198,136)
    rgb255=(243,194,132)
    rgb255=(243,192,130)
    rgb255=(244,189,126)
    rgb255=(244,186,124)
    rgb255=(244,183,122)
    rgb255=(244,180,120)
    rgb255=(244,176,118)
    rgb255=(244,173,116)
    rgb255=(245,170,115)
    rgb255=(245,167,113)
    rgb255=(245,164,111)
    rgb255=(245,160,109)
    rgb255=(245,156,107)
    rgb255=(245,154,107)
    rgb255=(244,150,106)
    rgb255=(244,147,106)
    rgb255=(244,144,105)
    rgb255=(244,140,104)
    rgb255=(244,137,104)
    rgb255=(243,133,103)
    rgb255=(243,130,103)
    rgb255=(243,126,102)
    rgb255=(242,124,103)
    rgb255=(241,120,103)
    rgb255=(240,116,104)
    rgb255=(239,113,104)
    rgb255=(238,109,105)
    rgb255=(238,107,105)
    rgb255=(237,102,105)
    rgb255=(236,98,106)
    rgb255=(235,95,106)
    rgb255=(233,92,108)
    rgb255=(232,89,109)
    rgb255=(230,85,110)
    rgb255=(229,83,111)
    rgb255=(227,79,112)
    rgb255=(225,75,113)
    rgb255=(224,72,114)
    rgb255=(222,68,116)
    rgb255=(220,65,116)
    rgb255=(217,62,118)
    rgb255=(214,58,120)
    rgb255=(212,56,121)
    rgb255=(209,53,123)
    rgb255=(207,50,124)
    rgb255=(204,47,126)
    rgb255=(201,43,128)
    rgb255=(199,40,129)
    rgb255=(196,37,131)
    rgb255=(193,36,132)
    rgb255=(189,34,134)
    rgb255=(186,33,135)
    rgb255=(181,31,137)
    rgb255=(177,29,139)
    rgb255=(174,28,141)
    rgb255=(169,26,143)
    rgb255=(166,25,144)
    rgb255=(161,24,146)
    rgb255=(155,25,147)
    rgb255=(151,25,149)
    rgb255=(145,26,150)
    rgb255=(141,27,151)
    rgb255=(135,28,153)
    rgb255=(130,29,154)
    rgb255=(124,29,156)
    rgb255=(117,30,158)
    rgb255=(112,31,159)
    rgb255=(103,33,160)
    rgb255=(98,34,161)
    rgb255=(88,36,162)
    rgb255=(78,37,163)
    rgb255=(70,38,164)
    rgb255=(57,39,165)
    rgb255=(46,40,165)
    rgb255=(24,41,166)
    rgb255=(0,42,167)
  },
  colormap/cetlinearworb/.style={colormap name=cetlinearworb,},
}

%% file: writing_utils.tex
\newcommand{\reva}[1]{#1}
\newcommand{\revb}[1]{#1}
\newcommand{\revc}[1]{#1}

%% file: sections/introduction.tex
\section{Introduction}\label{sec:introduction}

Living biological tissue on the microstructural scale consists of multiple constituents (e.g., collagen
fibers, elastin fibers, different kinds of cells, and more) with relevant effects on the macroscopic or organ
scale. Additionally, biological tissue can grow (change mass) and remodel (change microstructure)
continuously in response to mechanical and chemical cues with the ultimate goal of retaining a local
preferred mechanobiological environment, a so-called homeostatic state, or, better, to allow for allostasis,
an even more crucial process for living systems.

A key process involved in this mechanobiological activity is turnover, i.e., the continuous deposition of
new and degradation of existing mass \cite{Cyron2017a}. During tissue maintenance, turnover is balanced
such that there is no effective change in mass or microstructure. Unbalanced turnover can result in
excessive mass deposition, e.g., fibrosis after injury or disease.

Various computational models have been developed on different scales and complexity levels to increase
understanding of soft tissue growth and remodeling (G\&R). A family of methods that model these
key mechanobiological aspects (i.e., homeostasis and turnover) builds upon the constrained mixture
theory \cite{Humphrey2002a,Cyron2016a,Latorre2018a}.

The constrained mixture model \citep{Humphrey2002a} assumes tissue to consist of multiple constituents.
Mass increments of the constituents are added at every point in time into the mixture to model the
key mechanobiological process of production. Once deposited, the mass increments are actively degraded over time.
Models based upon the constrained mixture theory are widespread in
vascular G\&R \cite{Humphrey2021a} and were recently adopted to cardiac G\&R \cite{Gebauer2023a}.

Organ-scale constrained mixture models typically use the finite element method for spatial
discretization. Early models of thin-walled vascular geometries use 2D membrane elements
(e.g., \cite{Baek2006a,ZeinaliDavarani2011a,Sheidaei2011a})
and, later, also models with 3D elements have been developed \cite{Valentin2013a, Horvat2019a, Maes2023a}.
However, large organ-scale models are still missing, mainly due to the high computational costs.

In constrained mixture models, one needs to track the deformation state of every mass increment
deposited at every prior point in time, resulting in typically high computational costs.
Hence, different variants of the constrained
mixture theory have been developed to reduce computational costs and allow larger organ-scale
simulations.

There are so-called homogenized constrained mixture models \cite{Cyron2016a} that apply a temporal
homogenization of deposition and degradation of constituents. In these models, the integration of the
history variables reduces to ordinary differential equations. These models have been applied to
organ-scale models of vascular G\&R on 3D thick-walled cylinders \cite{Braeu2016a,Braeu2019a} and
patient-specific geometries \cite{Mousavi2019a}, and, recently, for cardiac G\&R \cite{Gebauer2023a}.
The homogenization can fail in situations with rapid changes in load and mass production, resulting in
appreciable differences between homogenized and classical constrained mixture models \reva{(see, for
    example, Section~3.2 and Figure~5 in \citet{Cyron2016a}).} \revb{Additionally, losing track of the
    individual mass increments precludes modeling increased degradation of individual fibers caused by excessive
    tension and damage as proposed by \citet{Baek2007a}.}

Another approach is to compute only the mechanobiological
steady-state at the end of active G\&R. This approach, called equilibrated constrained mixture model \cite{Latorre2018a,Latorre2020a},
is particularly useful if changes in the external loads are slow compared to the
involved turnover time constants. The equilibration can, however, not predict G\&R that does not
reach a new mechanobiological equilibrium, e.g., in cases where mechanobiological sensing is disrupted.
In this paper, we will analyze a mechanobiologically unstable G\&R example that cannot be described
by the equilibrated constrained mixture model.

Challenging G\&R scenarios, where both computationally cheaper variants fail, require to solve
the classical constrained mixture model.
In this work, we focus on the classical constrained mixture approach, where a large integral
over the deposition and degradation history is solved. These integrals are typically
solved with a trapezoidal rule \cite{ZeinaliDavarani2011a,Valentin2013a,Maes2023a} or
Simpson's rule \cite{Horvat2019a} with \emph{equal-sized} timesteps. To bound
computational costs, a few models assume a maximum survival time of deposited mass, after which contributions
to the current mixture response are neglected \cite{ZeinaliDavarani2011a,Horvat2019a,Maes2023a}.
This can lead to sudden jumps in behavior if the deformation since deposition is large and the
constituents have a non-linear material behavior.
Here, we propose to adaptively integrate history variables in constrained mixture models by exploiting
that the influence of deposited mass decreases over time through active degradation.
Adaptive history integration reduces the computational costs tremendously and still avoids the
definition of a maximum survival time while keeping the additional integration error negligibly small.
In our organ-scale example, we focus on cardiac G\&R, but the methods are applicable to any constrained
mixture model.

In Section~\ref{sec:modeling}, we briefly summarize the equations for constrained mixture models
in the context of finite element methods and describe two strategies to adaptively
integrate the G\&R history. Section~\ref{sec:results_zerod} compares the adaptive integration
approaches with the non-adaptive case on a simple tissue patch example. To demonstrate that the
adaptive integration enables constrained mixture models for large 3D organ-scale examples, we apply
the constrained mixture model to a patient-specific geometry of two ventricles in Section~\ref{sec:results_threed}.
Finally, we discuss and summarize the findings.

%% file: sections/modeling.tex
\section{Mathematical modeling}\label{sec:modeling}

To describe G\&R on the organ scale, we use nonlinear
continuum mechanics \cite{Holzapfel2000a}. The material point $\bs{X}$ in the reference configuration
$\mathcal{B}_0$ is mapped to its spatial point $\bs{x}$ in the current configuration
$\mathcal{B}_s$ at G\&R time $s$\footnote{We use $s$ for the G\&R time to emphasize that G\&R takes place on
    the large timescale that is different from the typical timescale of a heartbeat usually denoted with $t$.} via
\begin{align*}
    \bs{x}: \mathcal{B}_0 \times [0, \infty) \rightarrow \mathcal{B}_s, \quad (\bs{X}, s) \mapsto \bs{x}(\bs{X}, s).
\end{align*}
Without loss of generality, we assume that the configuration at $s=0$ is the reference configuration.
The displacement of point $\bs{X}$ is $\bs{u} = \bs{x} - \bs{X}$. The deformation gradient is
$\bs{F} = \frac{\pd \bs{x}}{\pd \bs{X}}$ with determinant $J=\det{\bs{F}}$.

The tissue consists of multiple structurally significant constituents, which are modeled individually.
These constituents are locally entangled and deform together, i.e., the displacement field of
each constituent is the same as the displacement field of the mixture ($\bs{u} = \bs{u}^i$). We denote quantities
related to a specific constituent with superscript $i$.

The stress response is homogenized across the constituents, i.e.,
\begin{align}
    \label{eqn:ruleofmixture}
    \bs{S} = 2\frac{\partial \Psi}{\partial \bs{C}} = \sum_i \rho_0^i 2\frac{\partial W^i}{\partial \bs{C}} = \sum_i \rho_0^i \tilde{\bs{S}}^i,
\end{align}
where $\bs{S}$ is second Piola-Kirchoff stress of the mixture, $\bs{C} = \bs{F}^\Trans \bs{F}$ is the right Cauchy-Green deformation tensor, $\Psi$ is the Helmholtz free-energy function of the
mixture, and for each constituent $i$, $\rho_0^i$ is the reference mass
density per mixture volume, $W^i$ is the strain energy per unit mass and  $\tilde{\bs{S}}^i$ is the fictitious, specific second Piola-Kirchhoff
stress tensor. We denote it \emph{fictitious} and \emph{specific} since it is the stress tensor of
the constituent only in case the mass fraction is one and after multiplication with the mass density. This homogenization across the constituents
of the stress response is called a simple \emph{rule-of-mixture} \citep{Humphrey2002a}
and has the benefit that only the equilibrium for the mixture has to be solved and not for
each constituent individually. The mechanical equilibrium written in
the form of the principle of virtual work \citep{Holzapfel2000a} is
\begin{align}
    \label{eqn:principle_of_virtual_work}
    \delta W = \frac{1}{2}\int_{\mathcal{B}_0} \bs{S} : \delta \bs{C} \, \dd V = 0.
\end{align}
We neglect inertial effects since G\&R is typically happening in the time scale of weeks and months.

\subsection{Mass production and removal}

We solve our equilibrium equations using a total Lagrangian approach. The production and degradation
of mass must, therefore, also be represented in the reference configuration. We describe the mass
change by an evolution of the reference mass density per mixture volume $\rho_0^i=\rho_0^i(s)$ \cite{Braeu2016a}.

The evolution of mass is driven by continuous deposition and degradation of tissue.
To model deposition and degradation individually, we split the net mass production
rate $\dot{\rho}_0^i$ into a true mass production rate $\dot{\rho}_{0+}^i \ge 0$ and true mass
degradation rate $\dot{\rho}_{0-}^i \le 0$, i.e.,
\begin{align}
    \label{eqn:split_net_mass_prod_rate}
    \dot{\rho}_0^i = \dot{\rho}_{0+}^i + \dot{\rho}_{0-}^i.
\end{align}
Both evolutions can be chosen individually with the constraint that deposition and degradation
must compensate during homeostasis, i.e., $\dot{\rho}_0^i=0$ for pure maintenance. To remove the dimensionality of equation~(\ref{eqn:split_net_mass_prod_rate}),
we scale the quantity by the initial reference mass density $\rho_0^i(s=0)$ and obtain the non-dimensional
growth scalar $\kappa^i$ with
\begin{align*}
    \kappa^i(s) = \frac{\rho_0^i(s)}{\rho_0^i(s=0)},
\end{align*}
which is one if the mass does not change, larger than one during growth, and smaller than
one during atrophy. Equation~(\ref{eqn:split_net_mass_prod_rate}) can then be written as
\begin{align*}
    \dot{\kappa}^i(s) = \dot{\kappa}_+^i + \dot{\kappa}_-^i, \quad \text{with} \quad \dot{\kappa}_\pm^i = \frac{\dot{\rho}_{0\pm}^i}{\rho_0^i(s=0)}.
\end{align*}

The kinetics of cell apoptosis and, generally, tissue degradation is complex and can
depend on the current mechanobiological environment (see, for example, \cite{Valentin2009a}). For simplicity, we
will assume a simple Poisson degradation process similar to radioactive decay. The dimensionless mass
removal rate is, therefore,
\begin{align}
    \label{eqn:poisson_degradation}
    \dot{\kappa}_-^i = -\frac{\kappa^i}{T^i},
\end{align}
where $T^i$ is the mean survival time of the constituent.

The survival function $q^{i(\tau)}$ is the fraction of the mass deposited at time $\tau$ that is still surviving
at time $s$. For a simple Poisson degradation, it is
\begin{align*}
    q^{i(\tau)}(s) = \exp(-\frac{s-\tau}{T^i}).
\end{align*}

To link new production of tissue to the mechanobiological state of the tissue,
we follow \citet{Braeu2016a} and assume that the true mass production rate is
\begin{align}
    \label{eqn:mass_production_rate}
    \dot{\kappa}_+^i = \kappa^i \left[\frac{1}{T^i} + \bs{K}_\sigma^i : \left(\bs{\sigma}_\text{R}^i - \bs{\sigma}_\text{h}^i\right)\right],
\end{align}
where $\bs{K}_\sigma^i$ is a gain-type second-order tensor.
The first term in brackets ensures that the basal mass production rate compensates for the continuous
degradation of mass with time constant $T^i$ during homeostasis (equation\,(\ref{eqn:poisson_degradation})). The second term is unequal to zero if the Cauchy stress differs
from the homeostatic stress, and hence, net mass will be produced or degraded for compensation. The
tensor $\bs{\sigma}_\text{R}^i$ is the co-rotated Cauchy stress of the constituent \cite{Braeu2016a}
and $\bs{\sigma}_\text{h}^i$ is its homeostatic setpoint.

Biological tissue often consists of fibrous constituents like muscle cells and collagen fibers.
These constituents are often modeled as quasi-one-dimensional fiber families. For such constituents,
only the components in the direction of the fiber are unequal to zero \cite{Braeu2016a}, such that equation (\ref{eqn:mass_production_rate})
reduces to
\begin{align}
    \label{eqn:mass_production_rate_1d}
    \dot{\kappa}_+^i = \kappa^i \left[\frac{1}{T^i} + k_\sigma^i \frac{\sigma^i-\sigma_\text{h}^i}{\sigma_\text{h}^i}\right],
\end{align}
with $k_\sigma^i = \sigma_\text{h}^i \bs{K}_\sigma^i : (\bs{f}_0^i \otimes \bs{f}_0^i)$,
$\sigma^i = \bs{\sigma}_\text{R}^i : (\bs{f}_0^i \otimes \bs{f}_0^i)$,
$\sigma_\text{h}^i = \bs{\sigma}_\text{h}^i : (\bs{f}_0^i \otimes \bs{f}_0^i)$,
and $\bs{f}_0^i$ is the unit vector pointing in the direction of the fiber in the reference configuration.

\subsection{Turnover}\label{sec:modeling:turnover}

\begin{figure}
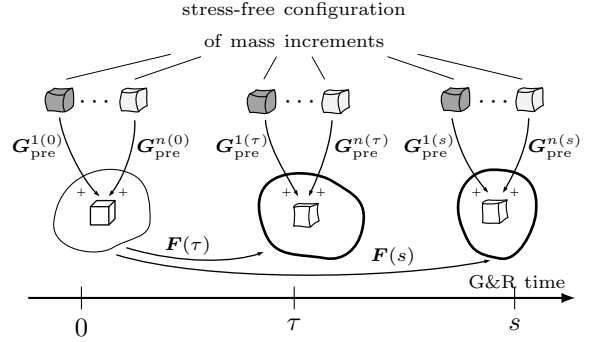

    \begin{center}
        \include{figures/constrained_mixture}
    \end{center}
    \caption{Configurations during G\&R. At every point in time, new mass increments of each constiuent
        are added to the mixture with a prestretch $\bs{G}_\text{pre}^{i(\tau)}$.}
    \label{fig:constrained_mixture}
\end{figure}

The constrained mixture theory \cite{Humphrey2002a} models turnover by continuously adding mass
increments into the mixture that will then degrade over time.

In every infinitesimal time interval $\dd s$, a new mass increment with mass $\dot{\rho}_{0+}^i \dd s$
and prestretch $\bs{G}_\text{pre}^{i}$ is added to the mixture as shown in Figure~\ref{fig:constrained_mixture}. Once deposited, the mass increment is
continuously degraded over time following the survival function $q^{i(\tau)}$. Hence, the reference mass
density evolves with
\begin{align*}
    \rho_0^i (s) = q^{i(0)}(s) \rho_0^i(s=0) + \int_0^s q^{i(\tau)}(s) \dot{\rho}_{0+}^i(\tau) \dd \tau.
\end{align*}
The first term outside of the integral describes the tissue that is initially present at $s=0$.
This equation can also be rewritten in terms of the nondimensional growth scalar, i.e.,
\begin{align}
    \label{eqn:current_growth_scalar}
    \kappa^i (s) = q^{i(0)}(s) + \int_0^s q^{i(\tau)}(s) \dot{\kappa}_+^i(\tau) \dd \tau.
\end{align}
Following the \emph{rule-of-mixture} approach of equation~(\ref{eqn:ruleofmixture}) on the constituent
level, the fictitious second Piola-Kirchhoff stress response of the
constituent $i$ is
\begin{align}
    \begin{aligned}
        \tilde{\bs{S}}^i (s) =\frac{1}{\kappa^i(s)}\Bigl[q^{i(0)}(s) \bs{\tilde{S}}^{i(0)}(\bs{F}_\text{e}^{i(0)}) + \cdots \\
            \int_0^s q^{i(\tau)}(s) \dot{\kappa}_+^i(\tau) \bs{\tilde{S}}^{i(\tau)}(\bs{F}_\text{e}^{i(\tau)}) \dd \tau\Bigr].
    \end{aligned}
    \label{eqn:current_constituent_stress_response}
\end{align}
where $\bs{\tilde{S}}^{i(\tau)}$ is the fictitious, specific second Piola-Kirchhoff stress of the mass increment added at
the intermediate time $\tau$ and $\bs{F}_\text{e}^{i(\tau)}$ is its total elastic deformation.
The total elastic deformation gradient for the initially present mass is
\begin{align*}
    \bs{F}_\text{e}^{i(0)} = \bs{F}(s) \bs{G}_\text{pre}^{i(0)},
\end{align*}
and for the later deposited mass, it is
\begin{align*}
    \bs{F}_\text{e}^{i(\tau)} = \bs{F}(s) \bs{F}^{-1}(\tau) \bs{G}_\text{pre}^{i(\tau)}.
\end{align*}
The fictitious, specific second Piola-Kirchhoff stress is usually computed from the strain energy per unit mass $W^i$ with
\begin{align*}
    \bs{\tilde{S}}^{i(\tau)} = 2 \frac{\pd W^i}{\pd \bs{F}_\text{e}^{i(\tau)}} : \frac{\pd \bs{F}_\text{e}^{i(\tau)}}{\pd \bs{C}}.
\end{align*}

\revb{It is often assumed} that the constituent is a fiber family that only creates stresses in one preferred
direction and that newly deposited fibers are always
aligned in the same direction \cite{Cyron2016a,Braeu2016a,Braeu2019a,Gebauer2023a,Mousavi2017a,Mousavi2019a}. \revb{In that case,} we can simplify
equation~(\ref{eqn:current_constituent_stress_response}) from
a tensor equation to the scalar equation
\begin{align*}
    \tilde{\sigma}^i (s) = \frac{1}{\kappa^i(s)}\Bigl[q^{i(0)}(s) \tilde{\sigma}^{i(0)}(\lambda_\text{e}^{i(0)}) + \cdots \\
    \int_0^s q^{i(\tau)}(s) \dot{\kappa}_+^i(\tau) \tilde{\sigma}^{i(\tau)}(\lambda_\text{e}^{i(\tau)}) \dd \tau\Bigr],
\end{align*}
where $\tilde{\sigma}^{i(\tau)}$ is the fictitious, specific fiber Cauchy stress, and
$\lambda_\text{e}^{i(\tau)}$ is the total \revc{elastic} stretch of the mass increment deposited at time $\tau$.
Initially, the total stretch of the fiber is $\lambda_\text{e}^{i(0)} = \lambda(s) \lambda_\text{pre}^{i(0)}$
and for later deposited increments, it is $\lambda_\text{e}^{i(\tau)} = \frac{\lambda(s) \lambda_\text{pre}^{i(\tau)}}{\lambda(\tau)}$,
where $\lambda_\text{pre}^{i(\tau)}$ is the deposition stretch (prestretch) of the fiber.
The fictitious, specific fiber Cauchy stress can be derived from the strain energy per unit mass via
\begin{align*}
    \tilde{\sigma}^{i(\tau)} = \lambda \frac{\pd W^i}{\pd \lambda_\text{e}^{i(\tau)}} \frac{\pd \lambda_\text{e}^{i(\tau)}}{\pd \lambda}
    = \frac{\lambda \lambda_\text{pre}^{i(\tau)}}{\lambda(\tau)} \frac{\pd W^i}{\pd \lambda_\text{e}^{i(\tau)}}.
\end{align*}
The fictitious, specific second Piola-Kirchhoff stress tensor can then be obtained with
\begin{align*}
    \tilde{\bs{S}}^i (s) = \frac{\tilde{\sigma}^i (s)}{\left(\lambda(s)\right)^2}\bs{f}_0^i \otimes \bs{f}_0^i.
\end{align*}
\revb{Note that we use the second Piola-Kirchhoff stress tensor here so that we can use the constant
    reference fiber direction $\bs{f}_0^i$ although the fibers rotate along the deformation.}

\revb{Also note that the following adaptive integration strategy does not necessitate a scalar contribution of the
    stresses but can handle any tensorial quantity.}

\subsection{Adaptive history integration}

The computationally expensive parts of constrained mixture models are the heredity integrals of the form
\begin{align}
    \label{eqn:form_of_integral}
    \int_0^s q^{i(\tau)} \bs{\mathcal{F}}(\tau, s) \dd \tau
\end{align}
needed for the growth scalar (equation (\ref{eqn:current_growth_scalar})) and the stress tensor
(equation (\ref{eqn:current_constituent_stress_response})). The possibly tensor-valued quantity $\bs{\mathcal{F}}$
represents tissue deposited into the mixture at the intermediate time $\tau$.
The integrand depends on the current deformation
of the mixture and, therefore, has to be re-evaluated at every timestep, i.e., cannot be stored
and reused from previous timesteps or Newton iterations.

The evaluation of function $\bs{\mathcal{F}}(\tau, s)$
requires information about the deformation state at intermediate times $\tau$. Hence, the function
can only be evaluated at points in time where the equilibrium~(\ref{eqn:principle_of_virtual_work}) has been
solved and stored. It is desirable to keep the number of stored configurations low to
minimize the needed memory consumption in large 3D organ-scale simulations.
Note that these are different needs than for classical adaptive quadrature rules, where one typically refines
intervals to reduce the numerical error of the computation.

\begin{figure}
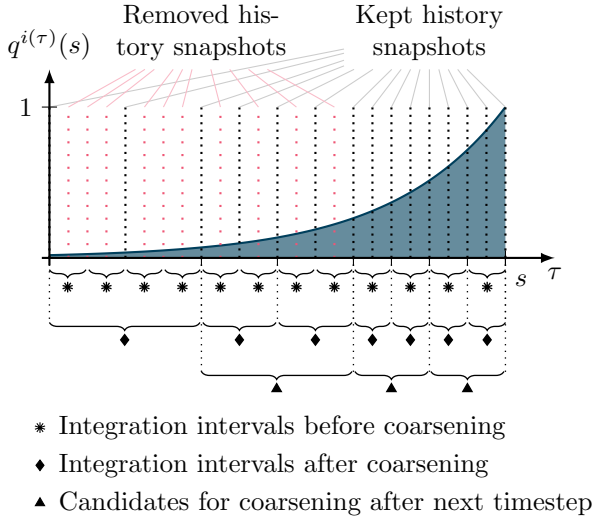

    \begin{center}
        \include{figures/adaptive_integration}
    \end{center}
    \caption{Fraction of tissue deposited at time $\tau$ surviving at time $s$ assuming a Poisson
        degradation process. With a non-adaptive G\&R history integration,
        the history snapshots at the red and black dotted lines need to be stored. Since the influence of
        at time $\tau$ deposited mass increments on the current stress state decreases in time, the integration
        timestep can be adaptively increased while keeping the integration error constant. Each
        integration rule consists of three equally spaced history snapshots assuming a Simpson integration rule.
        Each two subsequent Simpson intervals with equal timestep size (marked with *) are combined if the
        condition of the adaptive strategy is fulfilled.}
    \label{fig:adaptive_integration}
\end{figure}

The function $\bs{\mathcal{F}}$ can be discontinuous in time. These discontinuities are typically
imposed from the outside and known, e.g., discontinuous external loading. To keep the integration error low,
we divide the integration domain $[0,s]$ into intervals where $\bs{\mathcal{F}}$ is continuous.
Each interval is integrated by subsequently using the composed Newton-Cotes rule with three integration
points (Simpson's rule) \citep{Quarteroni2006a}. In the first timestep, where only two history snapshots are available, two integration
points are used (Trapezoidal rule) accordingly. We denote the numerical integration of function $f$
in the interval $\mathcal{I}$ using $n$ Newton-Cotes integration points with $\mathcal{Q}_n^\mathcal{I}(f)$.
The integration rules are given in appendix~\ref{app:newton_cotes}.

When using all history snapshots to integrate equation (\ref{eqn:form_of_integral}), the computational
costs grow linearly in time. With the adaptive integration, we exploit that tissue is degraded
over time following the survival function $q^{i(\tau)}$ (Figure~\ref{fig:adaptive_integration}).
The influence of deposited mass increments on the mixture response decreases over time.
The goal of the adaptive integration strategies is to keep the integration error constant over the
interval $[0,s]$.

We denote the integration error on the interval $\mathcal{I}$ of the numerical integration with
$E_\mathcal{Q}^\mathcal{I}$. Let us consider two consecutive Simpson intervals $\mathcal{I}_1$ and $\mathcal{I}_2$ with equal
length $\Delta s$, each consisting of three history snapshots (see pairs marked with * in Figure~\ref{fig:adaptive_integration}).
These two intervals are combined into a larger interval with length $2\Delta s$ if the numerical
integration error on the larger interval, $E_\mathcal{Q}^{\mathcal{I}_1+\mathcal{I}_2}$, is smaller
than a tolerance. To keep the integration error constant over each subinterval, the allowed tolerance
is scaled to the width of the combined interval, i.e., $\varepsilon_\mathcal{Q} \frac{2 \Delta s}{s}$,
where $\varepsilon_\mathcal{Q}$ is the allowed tolerance for the whole integral from $0$ to $s$. The
condition for coarsening is, therefore,
\begin{align}
    \label{eqn:adaptive_strategy}
    E_\mathcal{Q}^{\mathcal{I}_1+\mathcal{I}_2} \le \varepsilon_\mathcal{Q} \frac{2 \Delta s}{s}.
\end{align}
The interior point of both integration intervals $\mathcal{I}_1$ and $\mathcal{I}_2$ are then removed
from the history to free the memory. The resulting larger interval consists of three equally spaced
snapshots where Simpson's rule can be applied again.

The integration error $E_\mathcal{Q}^\mathcal{I}$ can, typically, not be determined analytically.
We propose two different strategies to approximate this error. The first strategy assumes that the basal
mass production rate is dominating G\&R such that an analytical expression of the error can be derived.
We call this strategy \emph{model equation} adaptive strategy.
The adaptive strategy is independent of the external loads such that the computational efforts can be
estimated ahead of the simulation. The second strategy uses a higher-order integration rule to
indicate the integration error with the benefit that the additional error can be kept small even in cases with
severe G\&R. We call it \emph{error indication} adaptive strategy.

\subsubsection{Strategy A: Model equation}

For the first strategy, we assume that the basal mass production in equation~(\ref{eqn:mass_production_rate}) is dominating G\&R.
In that case, the function $\bs{\mathcal{F}}(\tau, s)$ can be reduced to $\bs{\mathcal{F}}(\tau, s) = \frac{1}{T^i}$.
If we further assume that degradation follows a Poisson process with a constant rate, the survival
function $q^{i(\tau)}(s)$ can be pre-computed for each constituent, i.e.,
\begin{align*}
    q^{i(\tau)}(s) = q^i(s-\tau),
\end{align*}
where $q^i(\Delta s)$ is the fraction of mass still left after a timespan of $\Delta s$.

\begin{figure*}
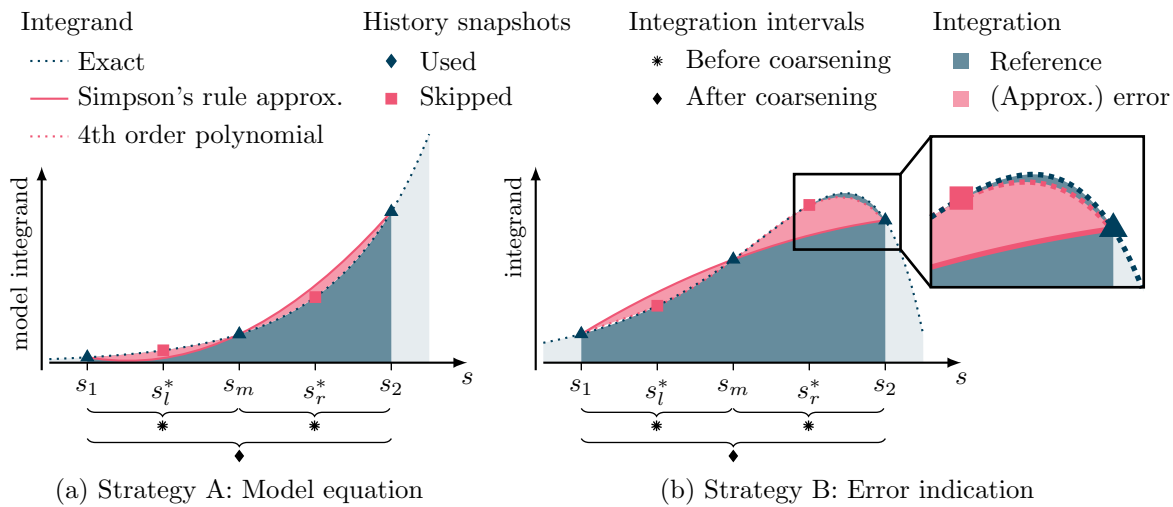

    \begin{center}
        \include{figures/adaptive_integration_interval}
    \end{center}
    \caption{Different strategies for coarsening the G\&R history. (a) The integrands occurring in
        constrained mixture models are approximated by the model equation that only captures the continuous tissue decay.
        This integrand can usually be integrated analytically, so the integration error is computed
        based on the analytical solution. (b) The integration error for the integrals occurring in
        constrained mixture models is approximated with a higher-order Newton-Cotes integration rule
        using five integration points (i.e., using a fourth-order polynomial approximation). If the
        integration error is below a given tolerance, the history items
        marked with $*$ are removed from the history, and the two Simpson intervals are combined.}
    \label{fig:adaptive_integration_interval}
\end{figure*}

For this special case, the integral on an arbitrary subinterval $\mathcal{I}=[s_1, s_2]$ is
\begin{align*}
    \int_\mathcal{I} q^i(s-\tau) \frac{1}{T^i} \dd \tau,
\end{align*}
and can be computed analytically:
\begin{align*}
    \int_{s_1}^{s_2} \frac{1}{T^i}\exp \left( -\frac{s-\tau}{T^i} \right) \dd \tau = \\ \left[
        \exp \left( -\frac{s-s_2}{T^i} \right)-\exp \left( -\frac{s-s_1}{T^i} \right)
        \right].
\end{align*}

The integration error of the model equation (see Figure~\ref{fig:adaptive_integration_interval} (a))
can then be expressed with
\begin{align*}
    E_\mathcal{Q}^\mathcal{I} = \left|\mathcal{Q}_3^\mathcal{I} (q^i)-\int_\mathcal{I} \frac{1}{T^i} q^i \dd \tau\right|.
\end{align*}
During pure maintenance (i.e., homeostasis), this exactly is the error of the integrals (\ref{eqn:current_growth_scalar})
and (\ref{eqn:current_constituent_stress_response}). However, the strategy does not see the amount of
net mass production. As a consequence, the strategy might be too aggressive in freeing history
snapshots in cases where net mass production dominates G\&R. We denote this strategy
with \emph{model equation} adaptive strategy.

\subsubsection{Strategy B: Error indication}

In the second strategy, we approximate the analytical integration with a higher-order integration scheme.
We have two subsequent Simpson intervals with the same timestep size $\Delta s$. Hence,
five quadrature points that are equally spaced can be used for integration, allowing to use
Boole's integration rule (Newton-Cotes with 5 quadrature points) as depicted in Figure~\ref{fig:adaptive_integration_interval} (b).
The approximated error is
\begin{align*}
    E_\mathcal{Q}(\mathcal{I}) = \left|\mathcal{Q}_3^\mathcal{I}(\mathcal{G})-\mathcal{Q}_5^\mathcal{I}(\mathcal{G})\right|,
\end{align*}
where $\mathcal{G}$ is the full integrand, i.e.,
\begin{align*}
    \mathcal{G} = q^{i(\tau)} \bs{\mathcal{F}}(\tau, s).
\end{align*}
This criterion does not rely on an analytical expression and can, therefore, be applied directly to the integrals
occurring in constrained mixture models, namely equation (\ref{eqn:current_growth_scalar}) for the
growth scalar and equation (\ref{eqn:current_constituent_stress_response}) for the Cauchy stress. The
integral for the Cauchy stress is normalized with the homeostatic Cauchy stress to result in a dimensionless
integral.
The interval is coarsened if both errors are below the prescribed integration tolerances $\varepsilon_\mathcal{Q}^\kappa$
and $\varepsilon_\mathcal{Q}^\sigma$. We denote this strategy
with \emph{error indication} adaptive strategy.

\subsection{Implementation}

\begin{algorithm}
    \include{figures/implementation_diagram}
    \caption{Implementation of a constrained mixture model with adaptive integration of history variables in a finite element framework.}
    \label{alg:implementation_diagram}
\end{algorithm}

To model organ-scale G\&R, the constrained mixture model is typically implemented in a finite element
framework on the integration point level as a constitutive model that computes the stress given the
deformation gradient. Algorithm~\ref{alg:implementation_diagram} summarizes the principal steps of
the model in such an environment.

In the element evaluation routine, the integrals of equations (\ref{eqn:current_growth_scalar}) and
(\ref{eqn:current_constituent_stress_response}) have to be solved using the composite Simpson's rule.
In case of a simple Poisson degradation process for the tissue, only the last snapshot in the integrals
depends on the current stress response of the constituent via
$\dot{\kappa}_+^i$ and equation (\ref{eqn:mass_production_rate}). Hence, to solve the integral, a
Newton-Raphson-type algorithm is applied locally. All older integration intervals can be integrated
ahead of the local Newton-Raphson algorithm (Step 5 in Algorithm~\ref{alg:implementation_diagram}) to
reduce computational costs. However, if tissue degradation depends on the current mechanobiological
environment, reintegrating over the whole history in every local Newton-Raphson step is necessary.
Once the growth scalar and the stress response are solved for each constituent,
the stress response of the mixture can be computed with equation (\ref{eqn:ruleofmixture}).

After the equilibrium equation has been solved, the current states of the mass increments need to
be added to the history. At this stage, older snapshots of the constrained mixture history can be
removed according to the adaptive strategy (Step 13 in Algorithm~\ref{alg:implementation_diagram}) to free memory and reduce the computational
costs of the following timesteps.

%% file: figures/constrained_mixture.tex
\begin{center}
    \begin{tikzpicture}
        \node[] (image) at (0,0) {\includegraphics{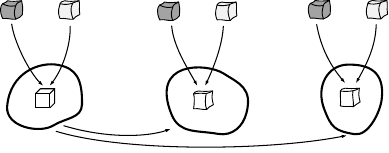}};

        \draw (-2.8,-0.1) node[scale=0.4] {$\bs{+}$};
        \draw (-2.25,-0.1) node[scale=0.4] {$\bs{+}$};

        \draw (-0.1,-0.1) node[scale=0.4] {$\bs{+}$};
        \draw (0.35,-0.1) node[scale=0.4] {$\bs{+}$};

        \draw (2.4,-0.1) node[scale=0.4] {$\bs{+}$};
        \draw (2.9,-0.1) node[scale=0.4] {$\bs{+}$};

        \draw[thick,-latex] (-3.5,-1.5) -- ++(7.2,0) node[above left] {\footnotesize{G\&R time}};
        \draw (-2.8,-1.35) -- ++(0,-0.3) node[below] {$\footnotesize{0}$};
        \draw (0,-1.35) -- ++(0,-0.3) node[below] {$\footnotesize{\tau}$};
        \draw (2.9,-1.35) -- ++(0,-0.3) node[below] {$\footnotesize{s}$};

        \draw (-2.6,1.1) node {$\cdots$};
        \draw (0.1,1.1) node {$\cdots$};
        \draw (2.65,1.1) node {$\cdots$};

        \draw (-1.4,-0.8) node{$\footnotesize{\bs{F}(\tau)}$};
        \draw (1.3,-0.95) node{$\footnotesize{\bs{F}(s)}$};

        \draw (-2.9,0.5) node[left] {$\footnotesize{\bs{G}_\text{pre}^{1(0)}}$};
        \draw (-2.2,0.5) node[right] {$\footnotesize{\bs{G}_\text{pre}^{n(0)}}$};

        \draw (-0.2,0.5) node[left] {$\footnotesize{\bs{G}_\text{pre}^{1(\tau)}}$};
        \draw (0.4,0.5) node[right] {$\footnotesize{\bs{G}_\text{pre}^{n(\tau)}}$};

        \draw (2.25,0.5) node[left] {$\footnotesize{\bs{G}_\text{pre}^{1(s)}}$};
        \draw (2.95,0.5) node[right] {$\footnotesize{\bs{G}_\text{pre}^{n(s)}}$};

        \draw (0,1.7) node[above,text width=3cm,align=center,name=incre_desc] {\footnotesize{stress-free configuration of mass increments}};

        \draw[line width=0.1pt] (-3,1.4) -- (incre_desc);
        \draw[line width=0.1pt] (-2.1,1.4) -- (incre_desc);

        \draw[line width=0.1pt] (-0.4,1.4) -- (incre_desc);
        \draw[line width=0.1pt] (0.4,1.4) -- (incre_desc);

        \draw[line width=0.1pt] (2.1,1.4) -- (incre_desc);
        \draw[line width=0.1pt] (3,1.4) -- (incre_desc);
    \end{tikzpicture}
\end{center}

%% file: figures/adaptive_integration.tex
\begin{tikzpicture}
    \fill[domain=0:6, samples=100, fill=dia4x1!60] (0,0) -- plot(\x,{2*exp(-(6-\x)/1.5)}) -- (6,0);
    \draw[domain=0:6, samples=100, dia4x1, thick] plot(\x,{2*exp(-(6-\x)/1.5)});
    \draw (6,0.1) -- ++(0,-0.2) node[below right] {$s$};

    \node[above, text width=2cm, align=center] (label_kept_history_points) at (5,2.5) {Kept history snapshots};
    \node[above, text width=3cm, align=center] (label_removed_history_points) at (2,2.5) {Removed history snapshots};

    \foreach \x in {0, 1, 2, 2.5, 3, 3.5, 4, 4.25, ..., 6}{
            \draw[thick,dotted] (\x,0) -- (\x,2);
            \draw[black!20] (\x,2) -- (label_kept_history_points);
        }
    \foreach \x in {0.25, 0.5, 0.75, 1.25, 1.5, 1.75, 2.25, 2.75, 3.25, 3.75}{
            \draw[thick,loosely dotted,dia4x3] (\x,0) -- (\x,2);
            \draw[dia4x3!40] (\x,2) -- (label_removed_history_points);
        }

    \foreach \x in {0, 2, 3, 4, 4.5, 5, 5.5, 6}{
            \draw[line width=0.5pt,dotted] (\x,0) -- (\x,-0.85);
        }
    \foreach \x in {2, 4, 5, 6}{
            \draw[line width=0.5pt,dotted] (\x,0) -- (\x,-1.55);
        }

    \draw[-latex, thick] (-0.1,0) -- (6.65,0) node[below]{$\tau$};
    \draw[-latex, thick] (0, -0.1) -- (0,2.5) node[above] {$q^{i(\tau)}(s)$};

    \draw (0.1, 2) -- ++(-0.2,0) node[left] {1};

    \foreach \x in {2,3,4,4.5,...,6}{
            \draw (\x,0.0) -- ++(0,-0.1);
        }

    \draw[thick,decorate,decoration={brace,amplitude=3pt}, transform canvas={yshift=-3pt}, line width=0.5pt] (0.5,-0.05)++(-0.02,0) -- node[below=3pt] {\pgfuseplotmark{10-pointed star}} ++(-0.48,0);
    \draw[thick,decorate,decoration={brace,amplitude=3pt}, transform canvas={yshift=-3pt}, line width=0.5pt] (1,-0.05)++(-0.02,0) -- node[below=3pt]   {\pgfuseplotmark{10-pointed star}} ++(-0.46,0);
    \draw[thick,decorate,decoration={brace,amplitude=3pt}, transform canvas={yshift=-3pt}, line width=0.5pt] (1.5,-0.05)++(-0.02,0) -- node[below=3pt] {\pgfuseplotmark{10-pointed star}} ++(-0.46,0);
    \draw[thick,decorate,decoration={brace,amplitude=3pt}, transform canvas={yshift=-3pt}, line width=0.5pt] (2,-0.05)++(-0.02,0) -- node[below=3pt]   {\pgfuseplotmark{10-pointed star}} ++(-0.46,0);
    \draw[thick,decorate,decoration={brace,amplitude=3pt}, transform canvas={yshift=-3pt}, line width=0.5pt] (2.5,-0.05)++(-0.02,0) -- node[below=3pt] {\pgfuseplotmark{10-pointed star}} ++(-0.46,0);
    \draw[thick,decorate,decoration={brace,amplitude=3pt}, transform canvas={yshift=-3pt}, line width=0.5pt] (3,-0.05)++(-0.02,0) -- node[below=3pt]   {\pgfuseplotmark{10-pointed star}} ++(-0.46,0);
    \draw[thick,decorate,decoration={brace,amplitude=3pt}, transform canvas={yshift=-3pt}, line width=0.5pt] (3.5,-0.05)++(-0.02,0) -- node[below=3pt] {\pgfuseplotmark{10-pointed star}} ++(-0.46,0);
    \draw[thick,decorate,decoration={brace,amplitude=3pt}, transform canvas={yshift=-3pt}, line width=0.5pt] (4,-0.05)++(-0.02,0) -- node[below=3pt]   {\pgfuseplotmark{10-pointed star}} ++(-0.46,0);
    \draw[thick,decorate,decoration={brace,amplitude=3pt}, transform canvas={yshift=-3pt}, line width=0.5pt] (4.5,-0.05)++(-0.02,0) -- node[below=3pt] {\pgfuseplotmark{10-pointed star}} ++(-0.46,0);
    \draw[thick,decorate,decoration={brace,amplitude=3pt}, transform canvas={yshift=-3pt}, line width=0.5pt] (5,-0.05)++(-0.02,0) -- node[below=3pt]   {\pgfuseplotmark{10-pointed star}} ++(-0.46,0);
    \draw[thick,decorate,decoration={brace,amplitude=3pt}, transform canvas={yshift=-3pt}, line width=0.5pt] (5.5,-0.05)++(-0.02,0) -- node[below=3pt] {\pgfuseplotmark{10-pointed star}} ++(-0.46,0);
    \draw[thick,decorate,decoration={brace,amplitude=3pt}, transform canvas={yshift=-3pt}, line width=0.5pt] (6,-0.05) -- node[below=3pt]              {\pgfuseplotmark{10-pointed star}} ++(-0.48,0);

    \draw[thick,decorate,decoration={brace,amplitude=3pt}, transform canvas={yshift=-3pt}, line width=0.5pt] (2,-0.75)++(-0.02,0) -- node[below=3pt]   {\pgfuseplotmark{diamond*}} ++(-1.98,0);
    \draw[thick,decorate,decoration={brace,amplitude=3pt}, transform canvas={yshift=-3pt}, line width=0.5pt] (3,-0.75)++(-0.02,0) -- node[below=3pt]   {\pgfuseplotmark{diamond*}} ++(-0.96,0);
    \draw[thick,decorate,decoration={brace,amplitude=3pt}, transform canvas={yshift=-3pt}, line width=0.5pt] (4,-0.75)++(-0.02,0) -- node[below=3pt]   {\pgfuseplotmark{diamond*}} ++(-0.96,0);
    \draw[thick,decorate,decoration={brace,amplitude=3pt}, transform canvas={yshift=-3pt}, line width=0.5pt] (4.5,-0.75)++(-0.02,0) -- node[below=3pt] {\pgfuseplotmark{diamond*}} ++(-0.46,0);
    \draw[thick,decorate,decoration={brace,amplitude=3pt}, transform canvas={yshift=-3pt}, line width=0.5pt] (5,-0.75)++(-0.02,0) -- node[below=3pt]   {\pgfuseplotmark{diamond*}} ++(-0.46,0);
    \draw[thick,decorate,decoration={brace,amplitude=3pt}, transform canvas={yshift=-3pt}, line width=0.5pt] (5.5,-0.75)++(-0.02,0) -- node[below=3pt] {\pgfuseplotmark{diamond*}} ++(-0.46,0);
    \draw[thick,decorate,decoration={brace,amplitude=3pt}, transform canvas={yshift=-3pt}, line width=0.5pt] (6,-0.75) -- node[below=3pt]              {\pgfuseplotmark{diamond*}} ++(-0.48,0);

    \draw[thick,decorate,decoration={brace,amplitude=3pt}, transform canvas={yshift=-3pt}, line width=0.5pt] (4,-1.4)++(-0.02,0) -- node[below=3pt] {\pgfuseplotmark{triangle*}} ++(-1.98,0);
    \draw[thick,decorate,decoration={brace,amplitude=3pt}, transform canvas={yshift=-3pt}, line width=0.5pt] (5,-1.4)++(-0.02,0) -- node[below=3pt] {\pgfuseplotmark{triangle*}} ++(-0.96,0);
    \draw[thick,decorate,decoration={brace,amplitude=3pt}, transform canvas={yshift=-3pt}, line width=0.5pt] (6,-1.4)++(-0.02,0) -- node[below=3pt] {\pgfuseplotmark{triangle*}} ++(-0.96,0);

    \draw (0,-2.25) node[left] {\pgfuseplotmark{10-pointed star}} node[right] {Integration intervals before coarsening};
    \draw (0,-2.75) node[left] {\pgfuseplotmark{diamond*}} node[right] {Integration intervals after coarsening};
    \draw (0,-3.25) node[left] {\pgfuseplotmark{triangle*}} node[right] {Candidates for coarsening after next timestep};
\end{tikzpicture}

%% file: figures/adaptive_integration_interval.tex
\begin{tikzpicture}
    \begin{scope}[xshift=0cm]
        \fill[domain=1.5:6.5, samples=100, fill=dia4x1!10] (1.5,0) -- plot(\x,{2*exp(-(6-\x)/1.2)}) -- (6.5,0);
        \fill[domain=2:6, samples=100, fill=dia4x1!60] (2,0) -- plot(\x,{2*exp(-(6-\x)/1.2)}) -- (6,0);

        \fill[domain=2:6, samples=20, dia4x3!60] plot(\x,{1.0808 + -0.8337*\x + 0.1645*\x^2}) -- plot(-\x+8,{2*exp(-(6-(-\x+8))/1.2)});

        \draw[domain=2:6, samples=20, dia4x3, thick] plot(\x,{1.0808 + -0.8337*\x + 0.1645*\x^2});

        \draw[domain=1.5:6.5, samples=100, dia4x1, thick, dotted] plot(\x,{2*exp(-(6-\x)/1.2)});

        \draw[-latex, thick] (1.5,0) -- (7,0) node[below] {$s$};

        \draw[-latex, thick] (1.4,-0.1) -- ++(0,2.7) node[above left,rotate=90] {\small model integrand};

        \draw (2,0) -- ++(0,-0.1) node[below] {$s_1$};
        \draw (6,0) -- ++(0,-0.1) node[below] {$s_2$};
        \draw (3,0) -- ++(0,-0.1) node[below] {$s_l^*$};
        \draw (4,0) -- ++(0,-0.1) node[below] {$s_m$};
        \draw (5,0) -- ++(0,-0.1) node[below] {$s_r^*$};

        \draw[dia4x1] (2,0.07134798669450479) node[scale=1.3] {\pgfuseplotmark{triangle*}};
        \draw[dia4x3] (3,0.1641699972477976) node[scale=1] {\pgfuseplotmark{square*}};
        \draw[dia4x1] (4,0.37775120567512366) node[scale=1.3] {\pgfuseplotmark{triangle*}};
        \draw[dia4x3] (5,0.8691964170141564) node[scale=1] {\pgfuseplotmark{square*}};
        \draw[dia4x1] (6,2.0) node[scale=1.3] {\pgfuseplotmark{triangle*}};

        \foreach \x in {2,3,4,5,6}{
                \draw (\x,0.0) -- ++(0,-0.1);
            }

        \draw[thick,decorate,decoration={brace,amplitude=3pt}, transform canvas={yshift=-3pt}, line width=0.5pt] (6,-0.5)++(-0.02,0) -- node[below=3pt] {\pgfuseplotmark{10-pointed star}} ++(-1.98,0);
        \draw[thick,decorate,decoration={brace,amplitude=3pt}, transform canvas={yshift=-3pt}, line width=0.5pt] (4,-0.5)++(-0.02,0) -- node[below=3pt] {\pgfuseplotmark{10-pointed star}} ++(-1.98,0);

        \draw[thick,decorate,decoration={brace,amplitude=3pt}, transform canvas={yshift=-3pt}, line width=0.5pt] (6,-0.9) -- node[below=3pt] {\pgfuseplotmark{diamond*}} ++(-4,0);

        \draw (4,-1.7) node{(a) Strategy A: Model equation};
    \end{scope}

    \begin{scope}[xshift=6.5cm]

        \fill[domain=1.5:6.5, samples=100, fill=dia4x1!10] (1.5,0) -- plot(\x,{(-0.02*(\x+0.5)^2+1)*exp(-(3-\x)/1.2)}) -- (6.5,0);
        \fill[domain=2:6, samples=100, fill=dia4x1!60] (2,0) -- plot(\x,{(-0.02*(\x+0.5)^2+1)*exp(-(3-\x)/1.2)}) -- (6,0);

        \fill[domain=2:6, samples=20, dia4x3!60] plot(\x,{-1.0781 + 0.8466*\x + -0.0587*\x^2}) -- plot(-\x+8,{-3.6376 + 4.9973*(-\x+8) + -2.3527*(-\x+8)^2 + 0.5044*(-\x+8)^3 + -0.0376*(-\x+8)^4});

        \draw[domain=2:6, samples=20, dia4x3, thick] plot(\x,{-1.0781 + 0.8466*\x + -0.0587*\x^2});

        \draw[domain=2:6, samples=20, dia4x3, thick, dotted] plot(\x,{-3.6376 + 4.9973*\x + -2.3527*\x^2 + 0.5044*\x^3 + -0.0376*\x^4});

        \draw[domain=1.5:6.5, samples=100, dia4x1, thick, dotted] plot(\x,{(-0.02*(\x+0.5)^2+1)*exp(-(3-\x)/1.2)});

        \draw[-latex, thick] (1.5,0) -- (7,0) node[below] {$s$};

        \draw[-latex, thick] (1.4,-0.1) -- ++(0,2.7) node[above left,rotate=90] {\small integrand};

        \draw (2,0) -- ++(0,-0.1) node[below] {$s_1$};
        \draw (6,0) -- ++(0,-0.1) node[below] {$s_2$};
        \draw (3,0) -- ++(0,-0.1) node[below] {$s_l^*$};
        \draw (4,0) -- ++(0,-0.1) node[below] {$s_m$};
        \draw (5,0) -- ++(0,-0.1) node[below] {$s_r^*$};


        \draw[dia4x1] (2.0,0.38027343244369344) node[scale=1.3] {\pgfuseplotmark{triangle*}};
        \draw[dia4x3] (3.0,0.755) node[scale=1.0] {\pgfuseplotmark{square*}};
        \draw[dia4x1] (4.0,1.369080655081231) node[scale=1.3] {\pgfuseplotmark{triangle*}};
        \draw[dia4x3] (5.0,2.091323569935662) node[scale=1.0] {\pgfuseplotmark{square*}};
        \draw[dia4x1] (6.0,1.8882865639090387) node[scale=1.3] {\pgfuseplotmark{triangle*}};

        \draw[line width=1pt] (4.8,1.5) -- ++(1.4,0) -- ++(0,1) -- ++(-1.4,0) -- cycle;
        \coordinate (zoom_small_top) at (6.2,2.5);
        \coordinate (zoom_small_bottom) at (6.2,1.5);

        \foreach \x in {2,3,4,5,6}{
                \draw (\x,0.0) -- ++(0,-0.1);
            }

        \draw[thick,decorate,decoration={brace,amplitude=3pt}, transform canvas={yshift=-3pt}, line width=0.5pt] (6,-0.5)++(-0.02,0) -- node[below=3pt] {\pgfuseplotmark{10-pointed star}} ++(-1.98,0);
        \draw[thick,decorate,decoration={brace,amplitude=3pt}, transform canvas={yshift=-3pt}, line width=0.5pt] (4,-0.5)++(-0.02,0) -- node[below=3pt] {\pgfuseplotmark{10-pointed star}} ++(-1.98,0);

        \draw[thick,decorate,decoration={brace,amplitude=3pt}, transform canvas={yshift=-3pt}, line width=0.5pt] (6,-0.9) -- node[below=3pt] {\pgfuseplotmark{diamond*}} ++(-4,0);

        \draw (5.5,-1.7) node{(b) Strategy B: Error indication};
    \end{scope}

    \begin{scope}[xshift=9.5cm,yshift=0cm]

        \begin{scope}[xshift=-6cm,yshift=-2cm,scale=2]
            \fill[domain=4.8:6.2, samples=30, fill=dia4x1!10] (4.8,1.5) -- plot(\x,{(-0.02*(\x+0.5)^2+1)*exp(-(3-\x)/1.2)}) -- (6.2,1.5);
            \fill[domain=4.8:6, samples=100, fill=dia4x1!60] (4.8,1.5) -- plot(\x,{(-0.02*(\x+0.5)^2+1)*exp(-(3-\x)/1.2)}) -- (6,1.5);

            \fill[domain=4.8:6, samples=20, dia4x3!60] plot(\x,{-1.0781 + 0.8466*\x + -0.0587*\x^2}) -- plot(-\x+10.8,{-3.6376 + 4.9973*(-\x+10.8) + -2.3527*(-\x+10.8)^2 + 0.5044*(-\x+10.8)^3 + -0.0376*(-\x+10.8)^4});

            \draw[domain=4.8:6, samples=20, dia4x3, line width=2pt] plot(\x,{-1.0781 + 0.8466*\x + -0.0587*\x^2});

            \draw[domain=4.8:6, samples=30, dia4x3, line width=2pt, dotted] plot(\x,{-3.6376 + 4.9973*\x + -2.3527*\x^2 + 0.5044*\x^3 + -0.0376*\x^4});

            \draw[domain=4.8:6.19, samples=100, dia4x1, line width=2pt, dotted] plot(\x,{(-0.02*(\x+0.5)^2+1)*exp(-(3-\x)/1.2)});

            \draw[dia4x3] (5.0,2.091323569935662) node[scale=2] {\pgfuseplotmark{square*}};
            \draw[dia4x1] (6.0,1.8882865639090387) node[scale=2.6] {\pgfuseplotmark{triangle*}};

            \draw[line width=1pt] (4.8,1.5) -- ++(1.4,0) -- ++(0,1) -- ++(-1.4,0) -- cycle;

            \draw[line width=1pt] (zoom_small_top) -- (4.8,2.5);
            \draw[line width=1pt] (zoom_small_bottom) -- (4.8,1.5);
        \end{scope}
    \end{scope}

    \begin{scope}[xshift=0cm,yshift=1cm]
        \draw(1,3.5) node[right] {Integrand};
        \draw[dia4x1, dotted, thick](1.25,3) -- ++(0.5,0);
        \draw (1.75,3) node[right] {Exact};

        \draw[dia4x3, thick](1.25,2.5) -- ++(0.5,0);
        \draw (1.75,2.5) node[right] {Simpson's rule approx.};

        \draw[dia4x3, dotted, thick](1.25,2) -- ++(0.5,0);
        \draw (1.75,2) node[right] {4th order polynomial};
    \end{scope}

    \begin{scope}[xshift=4.5cm,yshift=1cm]
        \draw(1,3.5) node[right] {History snapshots};
        \draw(1.5,3) node[dia4x1,scale=1.2] {\pgfuseplotmark{triangle*}};
        \draw(1.75,3) node[right] {Used};
        \draw(1.5,2.5) node[dia4x3,scale=1.0] {\pgfuseplotmark{square*}};
        \draw(1.75,2.5) node[right] {Skipped};
    \end{scope}

    \begin{scope}[xshift=8cm,yshift=1cm]
        \draw(1,3.5) node[right] {Integration intervals};
        \draw(1.5,3) node[] {\pgfuseplotmark{10-pointed star}};
        \draw(1.75,3) node[right] {Before coarsening};
        \draw(1.5,2.5) node[] {\pgfuseplotmark{diamond*}};
        \draw(1.75,2.5) node[right] {After coarsening};
    \end{scope}

    \begin{scope}[xshift=12cm,yshift=1cm]
        \draw(1,3.5) node[right] {Integration};
        \fill[dia4x1!60] (1.4,2.9) rectangle ++(0.2,0.2);
        \draw(1.75,3) node[right] {Reference};

        \fill[dia4x3!60] (1.4,2.4) rectangle ++(0.2,0.2);
        \draw(1.75,2.5) node[right] {(Approx.) error};
    \end{scope}
\end{tikzpicture}

%% file: figures/implementation_diagram.tex
\begin{center}
    \begin{tikzpicture}
        \newcommand*\deltax{0.3}
        \newcommand*\deltay{-0.5}
        \coordinate (timestepping) at (0,0);
        \draw (0,0) node[left] {\footnotesize{1}};
        \draw (timestepping) node[right] {Timestepping: While $s < s_\text{max}$};

        \coordinate (solve_equi) at (\deltax,\deltay);
        \draw (0,\deltay) node[left] {\footnotesize{2}};
        \draw (solve_equi) node[right] {Newton-Raphson-like \revb{algorithm} to solve (\ref{eqn:principle_of_virtual_work})};

        \coordinate (gploop) at (2*\deltax,2*\deltay);
        \draw (0,2*\deltay) node[left] {\footnotesize{3}};
        \draw (gploop) node[right, text width=7cm] {Iterate over elements and Gauss points};

        \coordinate (inner_gploop) at (3*\deltax,3*\deltay);
        \draw (0,3*\deltay) node[left] {\footnotesize{4}};
        \draw (inner_gploop) node[right] {Compute $\bs{F}$};

        \coordinate (inner_gploop2) at (3*\deltax,4*\deltay);
        \draw (0,4*\deltay) node[left] {\footnotesize{5}};
        \draw (inner_gploop2) node[right] {Solve $\int_0^{s-\Delta s} \cdots\,\dd s$ of (\ref{eqn:current_growth_scalar}) and (\ref{eqn:current_constituent_stress_response}) $\forall \, i$};

        \coordinate (inner_newton) at (3*\deltax,5*\deltay);
        \draw (0,5*\deltay) node[left] {\footnotesize{6}};
        \draw (inner_newton) node[right] {Newton-Raphson: Solve (\ref{eqn:current_growth_scalar}) and (\ref{eqn:current_constituent_stress_response}) $\forall \, i$};

        \coordinate (inner_newton_solve) at (4*\deltax,6.5*\deltay);
        \draw (0,6*\deltay) node[left] {\footnotesize{7}};
        \draw (inner_newton_solve) node[right, text width=8cm] {Compute residuum and linearization\\ of (\ref{eqn:current_growth_scalar}) and (\ref{eqn:current_constituent_stress_response}) };

        \coordinate (inner_newton_solve2) at (4*\deltax,8.5*\deltay);
        \draw (inner_newton_solve2) node[right, text width=8cm] {\emph{Note}: Reuse results from step 5\\ if possible};

        \coordinate (inner_newton_update) at (4*\deltax,10*\deltay);
        \draw (0,10*\deltay) node[left] {\footnotesize{8}};
        \draw (inner_newton_update) node[right, text width=7cm] {Newton step: Update $\kappa^i$ and $\tilde{\bs{S}}^i$};

        \coordinate (compute_s) at (3*\deltax,11*\deltay);
        \draw (0,11*\deltay) node[left] {\footnotesize{9}};
        \draw (compute_s) node[right, text width=7cm] {Compute $\bs{S}$ and linearization with (\ref{eqn:ruleofmixture})};

        \coordinate (compute_residuum) at (2*\deltax,12*\deltay);
        \draw (0,12*\deltay) node[left] {\footnotesize{10}};
        \draw (compute_residuum) node[right] {Compute residuum of (\ref{eqn:principle_of_virtual_work}) and linearization};

        \coordinate (update_disp) at (2*\deltax,13*\deltay);
        \draw (0,13*\deltay) node[left] {\footnotesize{11}};
        \draw (update_disp) node[right, text width=7cm] {Newton step: Update displacements};

        \coordinate (store_config) at (\deltax,14*\deltay);
        \draw (0,14*\deltay) node[left] {\footnotesize{12}};
        \draw (store_config) node[right, text width=7cm] {Store configuration of all mass increments};

        \coordinate (apply_adaptive_strategy) at (\deltax,15*\deltay);
        \draw (0,15*\deltay) node[left] {\footnotesize{13}};
        \draw (apply_adaptive_strategy) node[right, text width=7cm] {Apply adaptive strategy (\ref{eqn:adaptive_strategy})};

        \coordinate (update_time) at (\deltax,16*\deltay);
        \draw (0,16*\deltay) node[left] {\footnotesize{14}};
        \draw (update_time) node[right, text width=7cm] {Update G\&R time $s = s+\Delta s$};

        \draw[line width=1.5pt] (solve_equi)++(0,0.25) -- (update_time) -- ++(0,-0.2);
        \draw[line width=1.5pt] (gploop)++(0,0.25) -- (update_disp) -- ++(0,-0.2);
        \draw[line width=1.5pt] (inner_gploop)++(0,0.25) -- (compute_s) -- ++(0,-0.2);
        \draw[line width=1.5pt] (inner_newton_solve)++(0,0.4) -- (inner_newton_update) -- ++(0,-0.2);
    \end{tikzpicture}
\end{center}

%% file: sections/results_1D.tex
\section{Quasi-1D Tissue patch example}\label{sec:results_zerod}

\begin{figure}
    \import{figures/}{1d_tissue_patch.tex}
    \caption{The tissue patch with cross-sectional area $A$ consisting of one fiber
        family aligned in the longitudinal direction. The external loading and the support of the patch
        is such that the stress response is homogeneous in the patch.}
    \label{fig:tissue_geometry}
\end{figure}
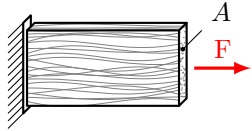

As a first example, we consider a quasi-1D tissue patch, as shown
in Figure~\ref{fig:tissue_geometry}. The tissue patch is loaded with an uniform traction with
resulting force $F$ in longitudinal direction. The support of the tissue patch is such that the stress response
is uniform within the patch, and the equilibrium equation can be written as
\begin{align}
    \label{eqn:1d_equilibrium}
    \sigma - \frac{F}{A} = 0,
\end{align}
where $A$ is the current cross-sectional area of the patch.

The tissue consists of a single family of Collagen fibers aligned in the longitudinal direction.
New fibers are deposited in the same direction as extant fibers with the homeostatic prestretch $\lambda_\text{h}$.
\revb{In case of unbalanced turnover, the tissue patch grows in cross-fiber direction such that the
    spatial mass density remains constant.}

The collagen fibers are modeled with a Fung-type strain energy function per unit mass, i.e.,
\begin{align}
    \label{eqn:fung_strain_energy}
    W = \frac{a}{2b} \left\{\exp \left[ b ({\lambda_\text{e}^{(\tau)}}^2-1)^2 \right]-1\right\},
\end{align}
where $a$ and $b$ are material parameters.
A Poisson degradation process, according to equation\,(\ref{eqn:poisson_degradation}), continuously degrades fibers
with the time constant $T$. Mass is produced according to equation\,(\ref{eqn:mass_production_rate_1d})
with growth gain factor $k_\sigma$.

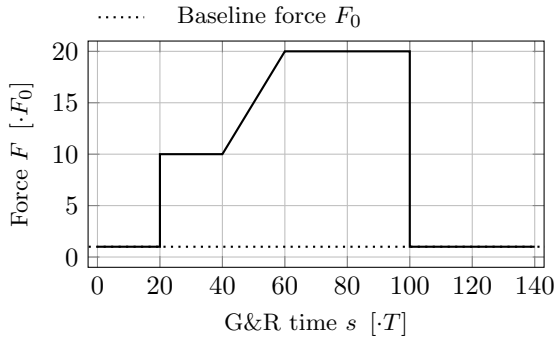
\begin{figure}
    \import{figures/}{1d_loading_scenario.tex}
    \caption{The loading scenario of the patch. $F_0$ is the initial force such that the patch is in
        homeostasis, and $T$ is the time-constant of the Poisson degradation process of the
        tissue.}
    \label{fig:tissue_loading_scenario}
\end{figure}

Initially, the tissue patch is loaded with $F_0$ such that the fibers, which are prestressed with $\lambda_\text{h}$,
are in homeostasis. The loading scenario is shown in Figure~\ref{fig:tissue_loading_scenario}.
This homeostatic load is held for $20 T$, where $T$ is the mean survival time of
the fiber. At $s=20 T$, the force is increased to $F=10F_0$ and held constant for $20 T$. Subsequently,
the force is linearly ramped up to $20F_0$ within $20T$ and kept constant for $40T$.
Finally, the baseline value $F_0$ of the force is restored and kept for $40T$.
This scenario ensures that the adaptive strategies are tested for different loading aspects that can
occur in-vivo, like homeostatic loading, a sudden and continuous load increase, and a reverse G\&R phase.

The parameters used for the simulations are given in Table\,\ref{tab:0d_parameters}. We integrate the
problem with a timestep size of $\Delta s = 0.75\,\si{days}$, which is a twentieth of the time constant $T$. We
compare the results using both adaptive strategies and tolerances of
$\varepsilon_\mathcal{Q}^\kappa=\varepsilon_\mathcal{Q}^\sigma=\varepsilon_\mathcal{Q} \in \{10^{-4}, 10^{-5}, \cdots,10^{-9}\}$
with non-adaptive history integration. The reference solution is obtained with a non-adaptive
history integration with ten times smaller timestep size.

\begin{table}
    \caption{Material parameters for the fibers of the tissue patch. Material parameters are taken from \citet{Braeu2016a}.}
    \label{tab:0d_parameters}
    \begin{tabular}{llll}
        \hline\noalign{\smallskip}
        Name                        & Parameter          & Value                        \\
        \noalign{\smallskip}\hline\noalign{\smallskip}
        \textit{Collagen parameters}                                                    \\
        Fung exponential parameters & $a$                & $568\,\si{\nicefrac{J}{kg}}$ \\
        ~                           & $b$                & $11.2$                       \\
        \textit{Homeostatic stretches}                                                  \\
        Homeostatic stretch         & $\lambda_\text{h}$ & $1.062$                      \\
        \textit{G\&R parameters}                                                        \\
        Mean survival time          & $T$                & $15\,\si{days}$              \\
        Growth gain                 & $k_\sigma$         & $\nicefrac{0.1}{T}$          \\
        \noalign{\smallskip}\hline
    \end{tabular}
\end{table}

\begin{figure*}
    \import{figures/}{1d_relative_error.tex}
    \caption{Relative Cauchy stress error, relative patch mass error, and the history size for the
        quasi-1D patch example integrated with the (a) model equation adaptive strategy and (b) error indication
        adaptive strategy and with a full integration. The reference solution is computed with full integration
        and a tenth of the timestep size.}
    \label{fig:tissue_relative_error}
\end{figure*}
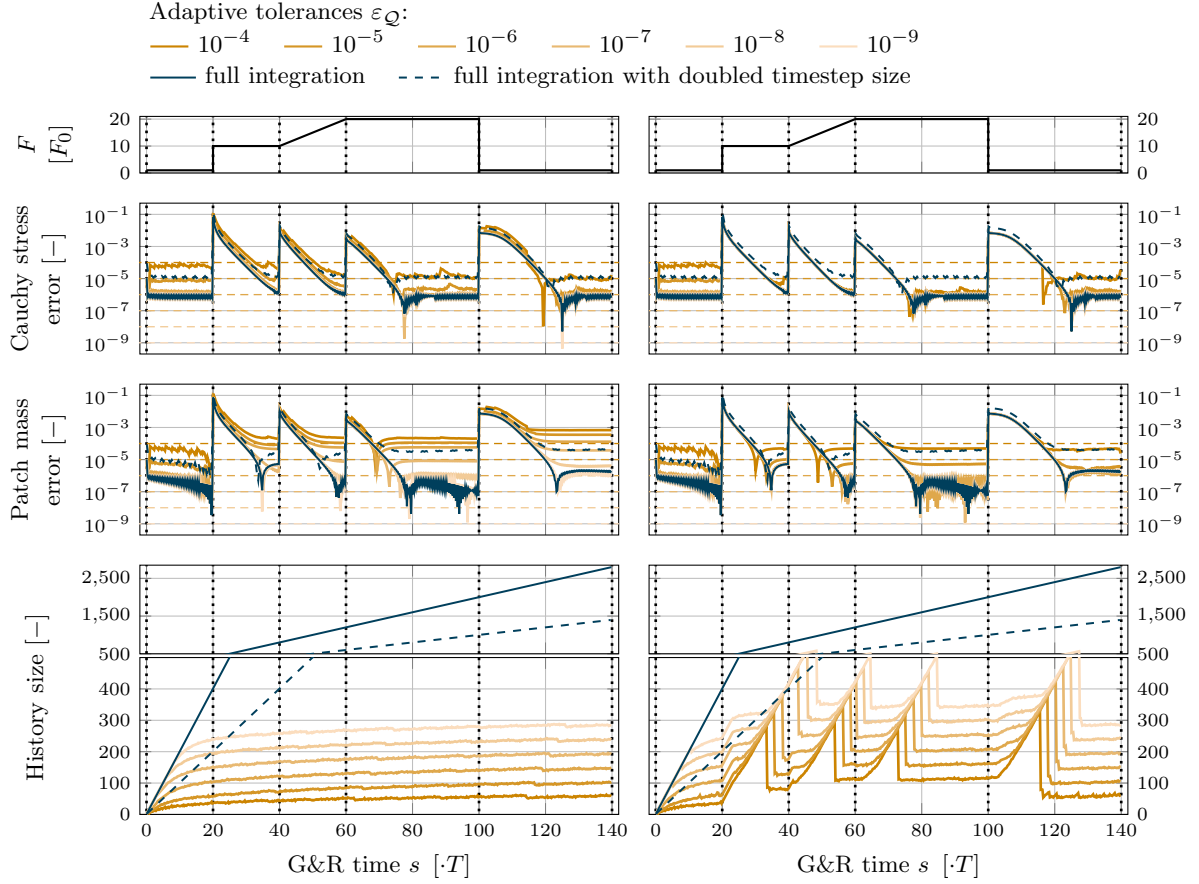

Figure~\ref{fig:tissue_relative_error} compares the two \revb{adaptive} strategies with different adaptive tolerances.
The first two rows (relative Cauchy stress error and relative patch mass error) result directly
from the two integrals occurring in constrained mixture models, i.e., equations
(\ref{eqn:current_constituent_stress_response}) and (\ref{eqn:current_growth_scalar}), respectively.

The relative errors can, naturally, not be smaller than for the full integration case. During the initial
maintenance phase, the relative error of the full integration is at around $10^{-6}$ for the Cauchy
stress and the patch mass. The errors of the adaptive integrations are on the same level except for
those cases with a higher adaptive tolerance. In the subsequent phases of G\&R, the relative errors
of the full integration and the error indication adaptive strategy are on the same level. The relative
error is higher for the model equation adaptive theory, especially when using loose adaptive tolerances.

It is important to note that the relative error of the Cauchy stress can recover the initial level after
a longer phase of constant load as the Cauchy stress converges (in mechanobiological stable G\&R)
towards the homeostatic stress. However, the relative patch mass error tends to accumulate over time as this
quantity is not a mechanobiologically controlled variable. This is especially pronounced for the
model equation adaptive strategy with loose adaptive tolerances, where the relative errors are larger.

The last row of Figure~\ref{fig:tissue_relative_error} shows the number of snapshots that are stored
in the history. The history size correlates directly with the computational costs of the model, i.e.,
the needed memory consumption and the evaluation time. The number of snapshots in the history for the model equation adaptive strategy
is independent of the load and can be computed ahead of time. When using the error
indication strategy, the history size increases after changes in the external load, resulting in changing memory demands
depending on the external loading. The peak history size for the
strictest analyzed tolerance ($\varepsilon_\mathcal{Q}=10^{-9}$) is below $300$ for the model equation
adaptive strategy and around $500$ for error indication adaptive strategy. Both strategies need
significantly fewer stored snapshots than the full integration with about $2800$ stored snapshots.
This difference is even more pronounced for looser adaptive tolerances. Since the history size
increases linearly for the full integration and plateaus for both adaptive strategies, the reduction
of computational costs will become even more drastic when advancing further in time.

In Figure~\ref{fig:tissue_relative_error}, we added a fully integrated case with a doubled timestep size for comparison (\revc{light blue}). The
total error is on the same level as the adaptive integration with the largest tolerance. However,
the final history size, which strongly correlates with evaluation time and memory consumption, is larger
by a factor of around $23$ with a tendency to increase further with the simulation time.

%% file: figures/1d_tissue_patch.tex
\begin{center}
    \begin{tikzpicture}
        \draw[thick,line cap=round,line join=round] (-0.05,-0.1) -- ++(0,1.2) -- ++(0.1,0.1) -- ++(0,-1.2) -- cycle;
        \foreach \x in {-0.1,0,0.1,...,1.2}
        \draw(-0.05,\x) -- ++(-0.2,-0.2);

        \fill[white] (0,0) -- ++(2,0) -- ++(0,1) -- ++(-2,0) -- cycle;
        \fill[white] (0,1) -- ++(0.1,0.1) -- ++(2,0) -- ++(-0.1,-0.1);
        \fill[white] (2,0) -- ++(0.1,0.1) -- ++(0,1);

        \draw[gray,very thin] (0,0.0603858477439598) to[out=-14.995000036765735,in=172.13082609818858] (2,0.07360147476447691);
        \draw[gray,very thin] (0,0.11413251675581892) to[out=-2.921510321681261,in=180.97002883103337] (2,0.11841834501061335);
        \draw[gray,very thin] (0,0.15923659605986332) to[out=5.534679219092482,in=164.69418982383849] (2,0.1522351929137758);
        \draw[gray,very thin] (0,0.1713678195303891) to[out=-8.040751464095841,in=193.01630482925538] (2,0.20917033961530462);
        \draw[gray,very thin] (0,0.23693697532458777) to[out=-0.02615128134379674,in=194.8288420012717] (2,0.2489177989556426);
        \draw[gray,very thin] (0,0.29358033991470456) to[out=7.765715671144861,in=197.0878644933968] (2,0.29277555107775993);
        \draw[gray,very thin] (0,0.35354003087164804) to[out=2.1934197806486324,in=189.26286508584454] (2,0.32922371907071946);
        \draw[gray,very thin] (0,0.3710879588060882) to[out=12.27334741587459,in=190.57887835146764] (2,0.4006978761462874);
        \draw[gray,very thin] (0,0.4125992144989861) to[out=-5.9510277576785064,in=165.0687744839668] (2,0.44853288995572543);
        \draw[gray,very thin] (0,0.455648498998714) to[out=15.498489801067784,in=198.0106976092978] (2,0.4711021203313104);
        \draw[gray,very thin] (0,0.5237235636122456) to[out=-11.65071604948003,in=191.01403646212208] (2,0.5059675973897455);
        \draw[gray,very thin] (0,0.5937425480179537) to[out=11.446478521274145,in=185.7916654141822] (2,0.5618548442519106);
        \draw[gray,very thin] (0,0.6261865449132354) to[out=-11.765892247806402,in=198.09177936155456] (2,0.6026586012247291);
        \draw[gray,very thin] (0,0.6527661657307168) to[out=12.295564982444922,in=196.3600276296881] (2,0.6713893450283599);
        \draw[gray,very thin] (0,0.6932654921800514) to[out=18.989344041439608,in=184.70848215044487] (2,0.7058728182661691);
        \draw[gray,very thin] (0,0.7589637359009697) to[out=1.7198773060004378,in=187.79067791129432] (2,0.7489918734342982);
        \draw[gray,very thin] (0,0.835288261535354) to[out=7.905879353275438,in=191.13834257484646] (2,0.8192403942497989);
        \draw[gray,very thin] (0,0.848231087458848) to[out=17.196458631558798,in=177.3321983816317] (2,0.8704168622162959);
        \draw[gray,very thin] (0,0.9080534470355786) to[out=2.459398846395011,in=171.39319328876377] (2,0.9073215185136123);
        \draw[gray,very thin] (0,0.9517076963653966) to[out=2.49203284315751,in=189.53378879347517] (2,0.9633060285035409);

        \draw[gray,very thin] (0.0037880356099341777,1.0037880356099342) to[out=-4.085834142733896,in=175.81810436549452] (2.0026695958458407,1.0026695958458405);
        \draw[gray,very thin] (0.030355705302990567,1.0303557053029906) to[out=-0.3794571361793313,in=179.30466758222482] (2.03173342756515,1.0317334275651497);
        \draw[gray,very thin] (0.05441840513300621,1.054418405133006) to[out=4.883789105454255,in=180.2993637493951] (2.0558855313695767,1.055885531369577);
        \fill[gray] (2.0227498770131365,0.595243562589268) circle (0.01);
        \fill[gray] (2.089653661635561,0.20636467469931805) circle (0.01);
        \fill[gray] (2.047447414113364,0.6265689546682199) circle (0.01);
        \fill[gray] (2.0145926014382725,0.2966463684943833) circle (0.01);
        \fill[gray] (2.071474772001871,0.12222723415872552) circle (0.01);
        \fill[gray] (2.0164117574911407,0.7548110745944097) circle (0.01);
        \fill[gray] (2.052396838313669,0.36976820603853977) circle (0.01);
        \fill[gray] (2.0435822779884845,0.5240088236019521) circle (0.01);
        \fill[gray] (2.0512513078531476,0.8035320482919543) circle (0.01);
        \fill[gray] (2.035935791698441,0.21079711934701983) circle (0.01);
        \fill[gray] (2.0512723128456893,0.16886046755743858) circle (0.01);
        \fill[gray] (2.0455253815028525,0.7800790406808976) circle (0.01);
        \fill[gray] (2.0697754622176294,0.2873008352468114) circle (0.01);
        \fill[gray] (2.0750189811040998,0.41681445676519624) circle (0.01);
        \fill[gray] (2.0413524497657005,0.23619491520329405) circle (0.01);
        \fill[gray] (2.050537280762493,0.641235582376789) circle (0.01);
        \fill[gray] (2.0885696290229387,1.0541706921616223) circle (0.01);
        \fill[gray] (2.0228224287600645,0.5858721502146019) circle (0.01);
        \fill[gray] (2.047883761183223,0.43658638103862435) circle (0.01);
        \fill[gray] (2.062305572829638,0.49695473190750933) circle (0.01);
        \fill[gray] (2.054052447863521,0.11963963799834237) circle (0.01);
        \fill[gray] (2.0179404167704185,0.2881592480386475) circle (0.01);
        \fill[gray] (2.063934999833814,0.9046792801892172) circle (0.01);
        \fill[gray] (2.027973136840071,0.3570042457432419) circle (0.01);
        \fill[gray] (2.065108780393327,1.041010131495713) circle (0.01);
        \fill[gray] (2.0686114872684556,0.39290200542322895) circle (0.01);
        \fill[gray] (2.076103041249933,0.9825660401723798) circle (0.01);
        \fill[gray] (2.0666663555779192,0.33314405625658505) circle (0.01);
        \fill[gray] (2.0477994856191897,0.12403106579936768) circle (0.01);
        \fill[gray] (2.025104424019268,0.8288230408808095) circle (0.01);
        \fill[gray] (2.026666072537487,0.6159215424051648) circle (0.01);
        \fill[gray] (2.034040592261155,0.8236791105776348) circle (0.01);
        \fill[gray] (2.0306410769321372,0.4718467403001288) circle (0.01);
        \fill[gray] (2.0221528710387577,0.15944807719951623) circle (0.01);
        \fill[gray] (2.041220864055444,0.6966057540449611) circle (0.01);
        \fill[gray] (2.0278341067146846,0.6243180619079833) circle (0.01);
        \fill[gray] (2.029450606705657,0.12395106383911034) circle (0.01);
        \fill[gray] (2.070772504617605,0.7460270903169892) circle (0.01);
        \fill[gray] (2.037405013317912,0.2767525216366682) circle (0.01);
        \fill[gray] (2.071446953677827,0.2715226538353368) circle (0.01);
        \fill[gray] (2.0574058656763454,0.4424351897466083) circle (0.01);
        \fill[gray] (2.0730404458248555,0.8372394724100045) circle (0.01);
        \fill[gray] (2.0135605573650275,0.9714162491876945) circle (0.01);
        \fill[gray] (2.0718405939708595,0.48118599969151055) circle (0.01);
        \fill[gray] (2.08918527156947,0.11688730737576937) circle (0.01);
        \fill[gray] (2.0738439823843318,0.17167528108732266) circle (0.01);
        \fill[gray] (2.0183652721729852,0.77329705906043) circle (0.01);
        \fill[gray] (2.062038597734625,0.34108196269355) circle (0.01);
        \fill[gray] (2.05156837398082,0.6916727484988179) circle (0.01);
        \fill[gray] (2.082330476398093,0.48115343520287346) circle (0.01);

        \draw[thick,line cap=round,line join=round] (0,0) -- ++(2,0) -- ++(0,1) -- ++(-2,0) -- cycle;
        \draw[thick,line cap=round,line join=round] (0,1) -- ++(0.1,0.1) -- ++(2,0) -- ++(-0.1,-0.1);
        \draw[thick,line cap=round,line join=round] (2,0) -- ++(0.1,0.1) -- ++(0,1);

        \draw[red,-latex, very thick] (2.2,0.5) -- node[above] {F} ++(0.75,0);

        \fill (2.05,0.75) circle(0.3mm);
        \draw (2.05,0.75) -- ++(0.25,0.25) node[above right] {$A$};



    \end{tikzpicture}
\end{center}

%% file: figures/1d_loading_scenario.tex
\begin{center}
    \begin{tikzpicture}
        \begin{axis}[
                scale only axis=true,
                width=6cm,
                height=3cm,
                xlabel=\small{G\&R time $s~\left[\cdot T\right]$},
                ylabel=\small{Force $F~\left[\cdot F_0\right]$},
                xmin=-3, xmax=143,
                ymin=-1, ymax=21,
                xtick={0,20,40,60,80,120,100,140},
                ytick={0,5,10,15,20},
                name=loading_scenario,
                grid=major
            ]

            \addplot[dotted,thick] coordinates{(-10,1) (144,1)};

            \addplot [thick] coordinates {
                    (0,1)    (20,1)    (20,10)   (40,10)
                    (60,20)  (100,20)  (100,1)  (140,1)
                };
        \end{axis}

        \node[anchor=south west,align=left] at (loading_scenario.north west) {%
            {\small
                    \begin{tabular}{@{}cl}
                        \begin{tikzpicture}
                            \fill[color=white] (0,-0.075) rectangle(0.6,0.15);
                            \draw[thick,dotted] (0,0) -- ++(0.6,0);
                        \end{tikzpicture} & Baseline force $F_0$
                    \end{tabular}
                }
        };
    \end{tikzpicture}
\end{center}

%% file: figures/1d_relative_error.tex
\begin{center}
    \begin{tikzpicture}
        \begin{axis}[
                ticklabel style = {font=\footnotesize},
                scale only axis=true,
                width=6.3cm,
                height=0.75cm,
                xlabel=,
                xticklabels=\empty,
                ylabel=\small{$F$ $[F_0]$},
                y label style={at={(axis description cs:-0.12,0.5)},text width=0.75cm,align=center},
                xmin=-2, xmax=142,
                ymin=0, ymax=21,
                ytick={0, 10, 20},
                grid=major,
                anchor=north,
                name=loading_scenario_model_equation,
            ]

            \addplot[color=black, line width=1pt, dotted] coordinates {(0,0) (0,21)};
            \addplot[color=black, line width=1pt, dotted] coordinates {(20,0) (20,21)};
            \addplot[color=black, line width=1pt, dotted] coordinates {(40,0) (40,21)};
            \addplot[color=black, line width=1pt, dotted] coordinates {(60,0) (60,21)};
            \addplot[color=black, line width=1pt, dotted] coordinates {(100,0) (100,21)};
            \addplot[color=black, line width=1pt, dotted] coordinates {(140,0) (140,21)};

            \addplot [thick] coordinates {
                    (0,1)    (20,1)    (20,10)   (40,10)
                    (60,20)  (100,20)  (100,1)  (140,1)
                };
        \end{axis}
        \begin{axis}[
                ticklabel style = {font=\footnotesize},
                scale only axis=true,
                width=6.3cm,
                height=2cm,
                xlabel=,
                xticklabels=\empty,
                ylabel=\small{Cauchy stress error $\left[-\right]$},
                y label style={text width=3cm,align=center},
                xmin=-2, xmax=142,
                ymin=2e-10, ymax=5e-1,
                ytick={1e-1, 1e-3, 1e-5, 1e-7, 1e-9},
                name=cauchy_stress_error_model_equation,
                grid=major,
                ymode=log,
                at={($(loading_scenario_model_equation.south)+(0,-0.4cm)$)},
                anchor=north,
            ]
            \addplot[color=black, line width=1pt, dotted] coordinates {(0,1e-13) (0,1e0)};
            \addplot[color=black, line width=1pt, dotted] coordinates {(20,1e-13) (20,1e0)};
            \addplot[color=black, line width=1pt, dotted] coordinates {(40,1e-13) (40,1e0)};
            \addplot[color=black, line width=1pt, dotted] coordinates {(60,1e-13) (60,1e0)};
            \addplot[color=black, line width=1pt, dotted] coordinates {(100,1e-13) (100,1e0)};
            \addplot[color=black, line width=1pt, dotted] coordinates {(140,1e-13) (140,1e0)};


            \addplot[color=dia6-3-1, line width=1pt] table [x=relative_time, y=model_equation_1e-4, col sep=comma] {figures/data/1d/relative_cauchy_stress_error.csv};
            \addplot[color=dia6-3-2, line width=1pt] table [x=relative_time, y=model_equation_1e-5, col sep=comma] {figures/data/1d/relative_cauchy_stress_error.csv};
            \addplot[color=dia6-3-3, line width=1pt] table [x=relative_time, y=model_equation_1e-6, col sep=comma] {figures/data/1d/relative_cauchy_stress_error.csv};
            \addplot[color=dia6-3-4, line width=1pt] table [x=relative_time, y=model_equation_1e-7, col sep=comma] {figures/data/1d/relative_cauchy_stress_error.csv};
            \addplot[color=dia6-3-5, line width=1pt] table [x=relative_time, y=model_equation_1e-8, col sep=comma] {figures/data/1d/relative_cauchy_stress_error.csv};
            \addplot[color=dia6-3-6, line width=1pt] table [x=relative_time, y=model_equation_1e-9, col sep=comma] {figures/data/1d/relative_cauchy_stress_error.csv};

            \addplot[color=dia1-2, line width=0.75pt] table [x=relative_time, y=full_integration, col sep=comma] {figures/data/1d/relative_cauchy_stress_error.csv};
            \addplot[color=dia1-3, line width=0.75pt] table [x=relative_time, y=full_integration_2, col sep=comma] {figures/data/1d/relative_cauchy_stress_error_coarse.csv};

        \end{axis}
        \begin{axis}[
                ticklabel style = {font=\footnotesize},
                scale only axis=true,
                width=6.3cm,
                height=2cm,
                xlabel=,
                xticklabels=\empty,
                ylabel=\small{Patch mass error $\left[-\right]$},
                y label style={text width=2.5cm,align=center},
                xmin=-2, xmax=142,
                ymin=2e-10, ymax=5e-1,
                ytick={1e-1, 1e-3, 1e-5, 1e-7, 1e-9},
                name=patch_mass_error_model_equation,
                at={($(cauchy_stress_error_model_equation.south)+(0,-0.4cm)$)},
                anchor=north,
                grid=major,
                ymode=log,
            ]
            \addplot[color=black, line width=1pt, dotted] coordinates {(0,1e-13) (0,1e0)};
            \addplot[color=black, line width=1pt, dotted] coordinates {(20,1e-13) (20,1e0)};
            \addplot[color=black, line width=1pt, dotted] coordinates {(40,1e-13) (40,1e0)};
            \addplot[color=black, line width=1pt, dotted] coordinates {(60,1e-13) (60,1e0)};
            \addplot[color=black, line width=1pt, dotted] coordinates {(100,1e-13) (100,1e0)};
            \addplot[color=black, line width=1pt, dotted] coordinates {(140,1e-13) (140,1e0)};


            \addplot[color=dia6-3-1, line width=1pt] table [x=relative_time, y=model_equation_1e-4, col sep=comma] {figures/data/1d/relative_growth_scalar_error.csv};
            \addplot[color=dia6-3-2, line width=1pt] table [x=relative_time, y=model_equation_1e-5, col sep=comma] {figures/data/1d/relative_growth_scalar_error.csv};
            \addplot[color=dia6-3-3, line width=1pt] table [x=relative_time, y=model_equation_1e-6, col sep=comma] {figures/data/1d/relative_growth_scalar_error.csv};
            \addplot[color=dia6-3-4, line width=1pt] table [x=relative_time, y=model_equation_1e-7, col sep=comma] {figures/data/1d/relative_growth_scalar_error.csv};
            \addplot[color=dia6-3-5, line width=1pt] table [x=relative_time, y=model_equation_1e-8, col sep=comma] {figures/data/1d/relative_growth_scalar_error.csv};
            \addplot[color=dia6-3-6, line width=1pt] table [x=relative_time, y=model_equation_1e-9, col sep=comma] {figures/data/1d/relative_growth_scalar_error.csv};

            \addplot[color=dia1-2, line width=0.75pt] table [x=relative_time, y=full_integration, col sep=comma] {figures/data/1d/relative_growth_scalar_error.csv};
            \addplot[color=dia1-3, line width=0.75pt] table [x=relative_time, y=full_integration_2, col sep=comma] {figures/data/1d/relative_growth_scalar_error_coarse.csv};
        \end{axis}
        \begin{axis}[
                ticklabel style = {font=\footnotesize},
                scale only axis=true,
                width=6.3cm,
                height=1.175cm,
                xlabel=,
                ylabel=,
                xmin=-2, xmax=142,
                ymin=500, ymax=2850,
                ytick={500, 1500, 2500},
                name=history_size_top_model_equation,
                at={($(patch_mass_error_model_equation.south)+(0,-0.4cm)$)},
                anchor=north,
                grid=major,
                xticklabels=\empty,
            ]
            \addplot[color=black, line width=1pt, dotted] coordinates {(0,500) (0,2850)};
            \addplot[color=black, line width=1pt, dotted] coordinates {(20,500) (20,2850)};
            \addplot[color=black, line width=1pt, dotted] coordinates {(40,500) (40,2850)};
            \addplot[color=black, line width=1pt, dotted] coordinates {(60,500) (60,2850)};
            \addplot[color=black, line width=1pt, dotted] coordinates {(100,500) (100,2850)};
            \addplot[color=black, line width=1pt, dotted] coordinates {(140,500) (140,2850)};

            \addplot[color=dia1-2, line width=0.75pt] table [x=relative_time, y=full_integration, col sep=comma] {figures/data/1d/history_size.csv};
            \addplot[color=dia1-3, line width=0.75pt] table [x=relative_time, y=full_integration_2, col sep=comma] {figures/data/1d/history_size_coarse.csv};
        \end{axis}
        \begin{axis}[
                ticklabel style = {font=\footnotesize},
                scale only axis=true,
                width=6.3cm,
                height=2.08cm,
                xlabel=\small{G\&R time $s~\left[\cdot T\right]$},
                ylabel=\small{History size $\left[-\right]$},
                y label style={at={(axis description cs:-0.17,0.75)},text width=3cm,align=center},
                xmin=-2, xmax=142,
                ymin=0, ymax=500,
                ytick={0, 100, 200, 300, 400},
                grid=major,
                anchor=north,
                name=history_size_bottom_model_equation,
                at={($(history_size_top_model_equation.south)+(0,-0.05cm)$)},
            ]
            \addplot[color=black, line width=1pt, dotted] coordinates {(0,500) (0,0)};
            \addplot[color=black, line width=1pt, dotted] coordinates {(20,500) (20,0)};
            \addplot[color=black, line width=1pt, dotted] coordinates {(40,500) (40,0)};
            \addplot[color=black, line width=1pt, dotted] coordinates {(60,500) (60,0)};
            \addplot[color=black, line width=1pt, dotted] coordinates {(100,500) (100,0)};
            \addplot[color=black, line width=1pt, dotted] coordinates {(140,500) (140,0)};

            \addplot[color=dia6-3-1, line width=1pt] table [x=relative_time, y=model_equation_1e-4, col sep=comma] {figures/data/1d/history_size.csv};
            \addplot[color=dia6-3-2, line width=1pt] table [x=relative_time, y=model_equation_1e-5, col sep=comma] {figures/data/1d/history_size.csv};
            \addplot[color=dia6-3-3, line width=1pt] table [x=relative_time, y=model_equation_1e-6, col sep=comma] {figures/data/1d/history_size.csv};
            \addplot[color=dia6-3-4, line width=1pt] table [x=relative_time, y=model_equation_1e-7, col sep=comma] {figures/data/1d/history_size.csv};
            \addplot[color=dia6-3-5, line width=1pt] table [x=relative_time, y=model_equation_1e-8, col sep=comma] {figures/data/1d/history_size.csv};
            \addplot[color=dia6-3-6, line width=1pt] table [x=relative_time, y=model_equation_1e-9, col sep=comma] {figures/data/1d/history_size.csv};

            \addplot[color=dia1-2, line width=0.75pt] table [x=relative_time, y=full_integration, col sep=comma] {figures/data/1d/history_size.csv};
            \addplot[color=dia1-3, line width=0.75pt] table [x=relative_time, y=full_integration_2, col sep=comma] {figures/data/1d/history_size_coarse.csv};
        \end{axis}

        \begin{axis}[
                ticklabel style = {font=\footnotesize},
                scale only axis=true,
                width=6.3cm,
                height=0.75cm,
                xlabel=,
                xticklabels=\empty,
                ylabel=,
                yticklabel pos=right,
                y label style={at={(axis description cs:-0.17,0.5)},text width=3cm,align=center},
                xmin=-2, xmax=142,
                ymin=0, ymax=21,
                ytick={0, 10, 20},
                grid=major,
                anchor=north,
                name=loading_scenario_higher_order,
                anchor=west,
                at={($(loading_scenario_model_equation.east)+(0.4cm,0)$)},
            ]

            \addplot[color=black, line width=1pt, dotted] coordinates {(0,0) (0,21)};
            \addplot[color=black, line width=1pt, dotted] coordinates {(20,0) (20,21)};
            \addplot[color=black, line width=1pt, dotted] coordinates {(40,0) (40,21)};
            \addplot[color=black, line width=1pt, dotted] coordinates {(60,0) (60,21)};
            \addplot[color=black, line width=1pt, dotted] coordinates {(100,0) (100,21)};
            \addplot[color=black, line width=1pt, dotted] coordinates {(140,0) (140,21)};

            \addplot [thick] coordinates {
                    (0,1)    (20,1)    (20,10)   (40,10)
                    (60,20)  (100,20)  (100,1)  (140,1)
                };
        \end{axis}
        \begin{axis}[
                ticklabel style = {font=\footnotesize},
                scale only axis=true,
                width=6.3cm,
                height=2cm,
                xlabel=,
                xticklabels=\empty,
                yticklabel pos=right,
                ylabel=,
                y label style={text width=3cm,align=center},
                xmin=-2, xmax=142,
                ymin=2e-10, ymax=5e-1,
                ytick={1e-1, 1e-3, 1e-5, 1e-7, 1e-9},
                name=cauchy_stress_error_error_estimation,
                at={($(loading_scenario_higher_order.south)+(0,-0.4cm)$)},
                grid=major,
                ymode=log,
                anchor=north,
            ]
            \addplot[color=black, line width=1pt, dotted] coordinates {(0,1e-13) (0,1e0)};
            \addplot[color=black, line width=1pt, dotted] coordinates {(20,1e-13) (20,1e0)};
            \addplot[color=black, line width=1pt, dotted] coordinates {(40,1e-13) (40,1e0)};
            \addplot[color=black, line width=1pt, dotted] coordinates {(60,1e-13) (60,1e0)};
            \addplot[color=black, line width=1pt, dotted] coordinates {(100,1e-13) (100,1e0)};
            \addplot[color=black, line width=1pt, dotted] coordinates {(140,1e-13) (140,1e0)};


            \addplot[color=dia6-3-1, line width=1pt] table [x=relative_time, y=higher_order_1e-4, col sep=comma] {figures/data/1d/relative_cauchy_stress_error.csv};
            \addplot[color=dia6-3-2, line width=1pt] table [x=relative_time, y=higher_order_1e-5, col sep=comma] {figures/data/1d/relative_cauchy_stress_error.csv};
            \addplot[color=dia6-3-3, line width=1pt] table [x=relative_time, y=higher_order_1e-6, col sep=comma] {figures/data/1d/relative_cauchy_stress_error.csv};
            \addplot[color=dia6-3-4, line width=1pt] table [x=relative_time, y=higher_order_1e-7, col sep=comma] {figures/data/1d/relative_cauchy_stress_error.csv};
            \addplot[color=dia6-3-5, line width=1pt] table [x=relative_time, y=higher_order_1e-8, col sep=comma] {figures/data/1d/relative_cauchy_stress_error.csv};
            \addplot[color=dia6-3-6, line width=1pt] table [x=relative_time, y=higher_order_1e-9, col sep=comma] {figures/data/1d/relative_cauchy_stress_error.csv};

            \addplot[color=dia1-2, line width=0.75pt] table [x=relative_time, y=full_integration, col sep=comma] {figures/data/1d/relative_cauchy_stress_error.csv};
            \addplot[color=dia1-3, line width=0.75pt] table [x=relative_time, y=full_integration_2, col sep=comma] {figures/data/1d/relative_cauchy_stress_error_coarse.csv};
        \end{axis}
        \begin{axis}[
                ticklabel style = {font=\footnotesize},
                scale only axis=true,
                width=6.3cm,
                height=2cm,
                xlabel=,
                xticklabels=\empty,
                yticklabel pos=right,
                ylabel=,
                y label style={text width=3cm,align=center},
                xmin=-2, xmax=142,
                ymin=2e-10, ymax=5e-1,
                ytick={1e-1, 1e-3, 1e-5, 1e-7, 1e-9},
                name=patch_mass_error_error_estimation,
                at={($(cauchy_stress_error_error_estimation.south)+(0,-0.4cm)$)},
                anchor=north,
                grid=major,
                ymode=log,
            ]
            \addplot[color=black, line width=1pt, dotted] coordinates {(0,1e-13) (0,1e0)};
            \addplot[color=black, line width=1pt, dotted] coordinates {(20,1e-13) (20,1e0)};
            \addplot[color=black, line width=1pt, dotted] coordinates {(40,1e-13) (40,1e0)};
            \addplot[color=black, line width=1pt, dotted] coordinates {(60,1e-13) (60,1e0)};
            \addplot[color=black, line width=1pt, dotted] coordinates {(100,1e-13) (100,1e0)};
            \addplot[color=black, line width=1pt, dotted] coordinates {(140,1e-13) (140,1e0)};


            \addplot[color=dia6-3-1, line width=1pt] table [x=relative_time, y=higher_order_1e-4, col sep=comma] {figures/data/1d/relative_growth_scalar_error.csv};
            \addplot[color=dia6-3-2, line width=1pt] table [x=relative_time, y=higher_order_1e-5, col sep=comma] {figures/data/1d/relative_growth_scalar_error.csv};
            \addplot[color=dia6-3-3, line width=1pt] table [x=relative_time, y=higher_order_1e-6, col sep=comma] {figures/data/1d/relative_growth_scalar_error.csv};
            \addplot[color=dia6-3-4, line width=1pt] table [x=relative_time, y=higher_order_1e-7, col sep=comma] {figures/data/1d/relative_growth_scalar_error.csv};
            \addplot[color=dia6-3-5, line width=1pt] table [x=relative_time, y=higher_order_1e-8, col sep=comma] {figures/data/1d/relative_growth_scalar_error.csv};
            \addplot[color=dia6-3-6, line width=1pt] table [x=relative_time, y=higher_order_1e-9, col sep=comma] {figures/data/1d/relative_growth_scalar_error.csv};

            \addplot[color=dia1-2, line width=0.75pt] table [x=relative_time, y=full_integration, col sep=comma] {figures/data/1d/relative_growth_scalar_error.csv};
            \addplot[color=dia1-3, line width=0.75pt] table [x=relative_time, y=full_integration_2, col sep=comma] {figures/data/1d/relative_growth_scalar_error_coarse.csv};
        \end{axis}
        \begin{axis}[
                ticklabel style = {font=\footnotesize},
                scale only axis=true,
                width=6.3cm,
                height=1.175cm,
                xlabel=,
                ylabel=,
                yticklabel pos=right,
                xmin=-2, xmax=142,
                ymin=500, ymax=2850,
                ytick={500, 1500, 2500},
                name=history_size_top_error_estimation,
                at={($(patch_mass_error_error_estimation.south)+(0,-0.4cm)$)},
                anchor=north,
                grid=major,
                xticklabels=\empty,
            ]
            \addplot[color=black, line width=1pt, dotted] coordinates {(0,500) (0,2850)};
            \addplot[color=black, line width=1pt, dotted] coordinates {(20,500) (20,2850)};
            \addplot[color=black, line width=1pt, dotted] coordinates {(40,500) (40,2850)};
            \addplot[color=black, line width=1pt, dotted] coordinates {(60,500) (60,2850)};
            \addplot[color=black, line width=1pt, dotted] coordinates {(100,500) (100,2850)};
            \addplot[color=black, line width=1pt, dotted] coordinates {(140,500) (140,2850)};
            \addplot[color=dia6-3-6, line width=1pt] table [x=relative_time, y=higher_order_1e-9, col sep=comma] {figures/data/1d/history_size.csv};

            \addplot[color=dia1-2, line width=0.75pt] table [x=relative_time, y=full_integration, col sep=comma] {figures/data/1d/history_size.csv};
            \addplot[color=dia1-3, line width=0.75pt] table [x=relative_time, y=full_integration_2, col sep=comma] {figures/data/1d/history_size_coarse.csv};

        \end{axis}
        \begin{axis}[
                ticklabel style = {font=\footnotesize},
                scale only axis=true,
                width=6.3cm,
                height=2.08cm,
                xlabel=\small{G\&R time $s~\left[\cdot T\right]$},
                yticklabel pos=right,
                ylabel=,
                y label style={at={(axis description cs:-0.17,0.75)},text width=3cm,align=center},
                xmin=-2, xmax=142,
                ymin=0, ymax=500,
                ytick={0, 100, 200, 300, 400},
                grid=major,
                anchor=north,
                name=history_size_bottom_error_estimation,
                at={($(history_size_top_error_estimation.south)+(0,-0.05cm)$)},
            ]
            \addplot[color=black, line width=1pt, dotted] coordinates {(0,500) (0,0)};
            \addplot[color=black, line width=1pt, dotted] coordinates {(20,500) (20,0)};
            \addplot[color=black, line width=1pt, dotted] coordinates {(40,500) (40,0)};
            \addplot[color=black, line width=1pt, dotted] coordinates {(60,500) (60,0)};
            \addplot[color=black, line width=1pt, dotted] coordinates {(100,500) (100,0)};
            \addplot[color=black, line width=1pt, dotted] coordinates {(140,500) (140,0)};

            \addplot[color=dia6-3-1, line width=1pt] table [x=relative_time, y=higher_order_1e-4, col sep=comma] {figures/data/1d/history_size.csv};
            \addplot[color=dia6-3-2, line width=1pt] table [x=relative_time, y=higher_order_1e-5, col sep=comma] {figures/data/1d/history_size.csv};
            \addplot[color=dia6-3-3, line width=1pt] table [x=relative_time, y=higher_order_1e-6, col sep=comma] {figures/data/1d/history_size.csv};
            \addplot[color=dia6-3-4, line width=1pt] table [x=relative_time, y=higher_order_1e-7, col sep=comma] {figures/data/1d/history_size.csv};
            \addplot[color=dia6-3-5, line width=1pt] table [x=relative_time, y=higher_order_1e-8, col sep=comma] {figures/data/1d/history_size.csv};
            \addplot[color=dia6-3-6, line width=1pt] table [x=relative_time, y=higher_order_1e-9, col sep=comma] {figures/data/1d/history_size.csv};

            \addplot[color=dia1-2, line width=0.75pt] table [x=relative_time, y=full_integration, col sep=comma] {figures/data/1d/history_size.csv};
            \addplot[color=dia1-3, line width=0.75pt] table [x=relative_time, y=full_integration_2, col sep=comma] {figures/data/1d/history_size_coarse.csv};
        \end{axis}

        \node[anchor=south west,align=left] at ($(loading_scenario_model_equation.north west)+(0,0.2cm)$) {%
        \small{Adaptive tolerances $\varepsilon_\mathcal{Q}$:}\\

        {\small
        \begin{tabular}{@{}c@{}lc@{}lc@{}lc@{}lc@{}lc@{}l}
            \begin{tikzpicture}
                \fill[color=white] (0,-0.075) rectangle(0.5,0.15);
                \draw[thick,color=dia6-3-1] (0,0) -- ++(0.5,0);
            \end{tikzpicture} & ~$10^{-4}$ &
            \begin{tikzpicture}
                \fill[color=white] (0,-0.075) rectangle(0.5,0.15);
                \draw[thick,color=dia6-3-2] (0,0) -- ++(0.5,0);
            \end{tikzpicture} & ~$10^{-5}$ &
            \begin{tikzpicture}
                \fill[color=white] (0,-0.075) rectangle(0.5,0.15);
                \draw[thick,color=dia6-3-3] (0,0) -- ++(0.5,0);
            \end{tikzpicture} & ~$10^{-6}$ &
            \begin{tikzpicture}
                \fill[color=white] (0,-0.075) rectangle(0.5,0.15);
                \draw[thick,color=dia6-3-4] (0,0) -- ++(0.5,0);
            \end{tikzpicture} & ~$10^{-7}$ &
            \begin{tikzpicture}
                \fill[color=white] (0,-0.075) rectangle(0.5,0.15);
                \draw[thick,color=dia6-3-5] (0,0) -- ++(0.5,0);
            \end{tikzpicture} & ~$10^{-8}$ &
            \begin{tikzpicture}
                \fill[color=white] (0,-0.075) rectangle(0.5,0.15);
                \draw[thick,color=dia6-3-6] (0,0) -- ++(0.5,0);
            \end{tikzpicture} & ~$10^{-9}$
        \end{tabular}
        }\\


        {\small
        \begin{tabular}{@{}c@{}lc@{}l}
            \begin{tikzpicture}
                \fill[color=white] (0,-0.075) rectangle(0.5,0.15);
                \draw[line width=0.75pt,color=dia1-2] (0,0) -- ++(0.5,0);
            \end{tikzpicture} & ~Full integration                            &
            \begin{tikzpicture}
                \fill[color=white] (0,-0.075) rectangle(0.5,0.15);
                \draw[line width=0.75pt,color=dia1-3] (0,0) -- ++(0.5,0);
            \end{tikzpicture} & ~Full integration with doubled timestep size
        \end{tabular}
        }\\

        {\small
        \begin{tabular}[t]{cl}
            \begin{tikzpicture}
                \draw[line width=1pt, densely dotted,black] (0.1,-0.2) -- ++(0,0.25);
            \end{tikzpicture} & Change of external loading
        \end{tabular}
        }
        };

        \draw ($(history_size_bottom_model_equation.south)+(0,-1.5)$) node{(a) Strategy A: Model equation};
        \draw ($(history_size_bottom_error_estimation.south)+(0,-1.5)$) node{(b) Strategy B: Error indication};
    \end{tikzpicture}
\end{center}

%% file: sections/results_3D.tex
\section{3D cardiac example}\label{sec:results_threed}

To demonstrate that the adaptive integration enables organ-scale simulations of constrained mixture models, we apply
it to a patient-specific model of the two ventricles. The geometry is segmented from MRI data
of a healthy female subject. Following \citet{Gebauer2023a}, we use the data from the
end-systolic configuration. We use a mesh with 67412 second-order tetrahedral elements generated
with Gmsh \cite{Geuzaine2009a} depicted in Figure~\ref{fig:3d_geometry_and_bc}.
We use a Newton algorithm with backtracking for solving the nonlinear
equilibrium equations and a conjugate gradient method for linear systems \citep{trilinos-website}
implemented in our in-house research code 4C \cite{4C} written in C++.

\subsection{Modeling}\label{sec:resultsthreed:modeling}

\begin{figure}
    \import{figures/}{3d_geometry_and_bc.tex}
    \caption{The patient-specific model of ventricles of a healthy female subject. The color
        represents the surfaces where different boundary conditions are applied.}
    \label{fig:3d_geometry_and_bc}
\end{figure}
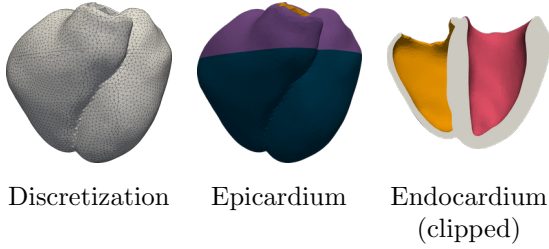

We apply different boundary conditions on the surfaces defined in Figure~\ref{fig:3d_geometry_and_bc}.
The influence of the pericardial sac on the epicardium is modeled with
a Robin boundary condition \cite{Pfaller2019a} in reference surface normal direction on the surface $\Gamma_\text{epi}$ \revb{with linear spring stiffness $c_\text{p}$}.
On the basal surface $\Gamma_\text{b}$, omnidirectional Robin boundary conditions \revb{with linear spring stiffness $c_\text{b}$} mimic the support
of the atria and adipose tissue. The blood in both ventricles is modeled with a pressure boundary condition of the endocardia of the left- and right ventricle
($\Gamma_\text{lv}$ and $\Gamma_\text{rv}$) \revb{with blood pressure values $p_\text{lv}$ and $p_\text{rv}$, respectively}. The resulting virtual work formulation is
\begin{align*}
    \delta W & = \frac{1}{2}\int_{\mathcal{B}_0} \bs{S}: \delta \bs{C} \, \dd V                          \\
             & + \int_{\Gamma_\text{epi}} (c_\text{p} \bs{u} \cdot \bs{N}_0) \delta \bs{u} \, \dd \Gamma
    + \int_{\Gamma_\text{b}} c_\text{b} \bs{u} \delta \bs{u} \, \dd \Gamma                               \\
             & + \int_{\Gamma_\text{lv}} p_\text{lv} J \bs{F}^{-\Trans} \bs{N}_0 \, \dd \Gamma
    + \int_{\Gamma_\text{rv}} p_\text{rv} J \bs{F}^{-\Trans} \bs{N}_0  \, \dd \Gamma                     \\
             & = 0,
\end{align*}
where $\bs{N}_0$ is the outward reference surface normal.

We follow \citet{Gebauer2023a} for modeling the constituents of the myocardium. That are cardiomyocytes
($i=\text{m}$) and 4 Collagen fiber families ($i=\text{c}$), all modeled as quasi-1D constituents and
computed for the patient-specific geometry with a rule-based algorithm \cite{Bayer2012a} by
prescribing the helix angle $\varphi$ at the endo- and epicardium.
The remaining constituents (mainly elastin) are combined in an isotropic constituent ($i=3D$). During
adulthood, no functional elastin is deposited into the matrix. Hence, we assume that the isotropic
constituent does not turnover ($k^{3D}=0$, $T^{3D}\rightarrow\infty$).

The strain energy per unit mass of collagen fibers
in the direction $\bs{f}^{\text{c}_i}$ are modeled with a Fung exponential strain energy function,
analogously to equation~(\ref{eqn:fung_strain_energy}), with material parameters $a^\text{c}$ and $b^\text{c}$.

The strain energy function of cardiomyocytes additively split\revb{s} into a passive and active contribution:
\begin{align*}
    W^\text{m} = W_\text{pas}^\text{m} + W_\text{act}^\text{m}.
\end{align*}
As with the collagen fibers, the strain energy function per unit mass for the passive response of
cardiomyocytes is a Fung exponential strain energy function with material parameters $a^\text{m}$ and $b^\text{m}$.
The active contribution is
\begin{align*}
    W_\text{act}^\text{m} = \frac{\sigma_\text{act}^\text{m}}{\rho_0{s=0}} \biggl[ \lambda_\text{act}^\text{m} + \frac{1}{3} \frac{(\lambda_\text{max}^\text{m}-\lambda_\text{act}^\text{m})^3}{(\lambda_\text{0}^\text{m})^2} \biggr],
\end{align*}
with $\frac{\pd \lambda_\text{act}^\text{m}}{\pd \lambda^{\text{m}(\tau)}} = \frac{1}{\lambda^{\text{m}(\tau)}}$ \citep{Braeu2019a,Wilson2012a}.

The strain energy function per unit mass of the remaining constituents is an isotropic neo-Hookean
strain energy contribution of the form
\begin{align*}
    W^\text{3D} = c_1 (\bar{I}_1-3),
\end{align*}
where $\bar{I}_1$ is the first modified invariant \citep{Holzapfel2000a} of the total elastic Cauchy-Green
deformation tensor with the isotropic and rotation-free prestretch tensor $\bs{G}^\text{3D}$. The latter
maps the stress-free configuration of elastin into the reference configuration.

We follow \citet{Braeu2019a} to model anisotropic growth of the tissue. They describe growth
by an elastic swelling of all existing constituents with a modified penalty formulation. The
additional term of the strain energy function is
\begin{align*}
    \Psi^\# = \frac{\kappa}{2} \bigl( |\bs{F}|-\frac{\rho_0(s)}{\rho_0(s=0)} \bigr)^2,
\end{align*}
where $\kappa$ is a penalty parameter. A sufficiently high penalty parameter ensures a constant
spatial density of the mixture. \revc{The resulting} tissue growth is anisotropic and happens mainly in the
direction of the smallest stiffness. \revb{A key advantage of that theory is that the anisotropy of
    G\&R naturally emerges from the anisotropic stiffness of the tissue and does not rely on a phenomenological
    definition of a growth tensor. Note that our adaptive integration strategies can also be
    combined with inelastic growth laws.}

\subsection{Loading scenario}\label{sec:resultsthreed:loading}

The configuration obtained from MRI data is not stress-free in the depicted configuration. Since
the data is from a healthy subject, we assume the configuration is in homeostasis, i.e., all
fiber constituents are in their preferred mechanical state. The baseline blood pressure in the left and right
ventricle is given by $p_\text{lv}$ and $p_\text{rv}$, respectively. We apply the prestress algorithm presented
in \citet{Gebauer2023a} until the maximum Euclidean norm of the nodal displacements falls below
$\varepsilon_\text{pre}$ to ensure that the initial configuration is close to the reference (imaged)
configuration.
The subsequent phase of G\&R of $200\,\si{days}$ with blood pressure levels remaining at baseline
ensure mechanobiological equilibrium of all constituents.

We simulate two hypertension conditions by elevating the left ventricular pressure at $s=0$ to
$p_\text{lv}^+=140\,\si{mmHg}$ and $p_\text{lv}^+=180\,\si{mmHg}$. The resulting G\&R are
mechanobiologically stable and unstable, respectively.

\begin{table}
    \caption{Material and Simulation parameters used for the 3D organ-scale example. The parameters
        are adopted from \citet{Braeu2016a} and \citet{Gebauer2023a}.}
    \label{tab:parameters}
    \begin{tabular}{llll}
        \hline\noalign{\smallskip}
        Name                           & Param.                          & Value                                \\
        \noalign{\smallskip}\hline\noalign{\smallskip}
        \textit{Fibers}                                                                                         \\
        ~Epicardial fiber helix angle  & $\varphi_\text{epi}$            & $-60^\circ$                          \\
        ~Endocardial fiber helix angle & $\varphi_\text{endo}$           & $60^\circ$                           \\
        \textit{Material parameters}                                                                            \\
        Myocytes:                                                                                               \\
        ~Fung exponential parameters   & $a^\text{m}$                    & $7.6\,\nicefrac{\si{J}}{\si{kg}}$    \\
        ~                              & $b^\text{m}$                    & $11.4$                               \\
        ~Active muscle tone            & $\sigma_{\text{act}}^\text{m}$  & $54\,\si{kPa}$                       \\
        ~                              & $\lambda_{0}^\text{m}$          & $0.8$                                \\
        ~                              & $\lambda_{\text{max}}^\text{m}$ & $1.4$                                \\
        Collagen:                                                                                               \\
        ~Fung exponential parameters   & $a^\text{c}$                    & $568\,\nicefrac{\si{J}}{\si{kg}}$    \\
        ~                              & $b^\text{c}$                    & $11.2$                               \\
        Elastin matrix:                                                                                         \\
        ~Neohookean parameter          & $a^\text{e}$                    & $72\,\nicefrac{\si{J}}{\si{kg}}$     \\
        Volumetric penalty             & $\kappa$                        & $150\,\si{kPa}$                      \\
        \textit{Boundary conditions}                                                                            \\
        Spring stiffness:                                                                                       \\
        ~Base                          & $c_\text{base}$                 & $2.0\,\nicefrac{\si{Pa}}{\si{mm}}$   \\
        ~Pericardium                   & $c_\text{p}$                    & $0.2\,\nicefrac{\si{Pa}}{\si{mm}}$   \\
        Baseline pressure:                                                                                      \\
        ~Left ventricle                & $p_\text{lv}$                   & $120\,\si{mmHg}$                     \\
        ~Right ventricle               & $p_\text{rv}$                   & $24\,\si{mmHg}$                      \\
        \multicolumn{3}{l}{Elevated left ventricular pressure:}                                                 \\
        ~Stable G\&R                   & $p_\text{lv}^+$                 & $140\,\si{mmHg}$                     \\
        ~Unstable G\&R                 & $p_\text{lv}^+$                 & $180\,\si{mmHg}$                     \\
        \textit{Initial conditions}                                                                             \\
        ~Reference mass density        & $\rho_0$                        & $1050\,\nicefrac{\si{kg}}{\si{m^3}}$ \\
        ~Myocyte mass fraction         & $\xi^\text{m}$                  & $0.6$                                \\
        ~Collagen mass fraction        & $\xi^\text{c}$                  & $0.1$                                \\
        ~Elastin mass fraction         & $\xi^\text{e}$                  & $0.3$                                \\
        \textit{G\&R parameters}                                                                                \\
        Myocytes:                                                                                               \\
        ~Homeostatic stretch           & $\lambda_\text{h}^\text{m}$     & $1.1$                                \\
        ~Growth gain                   & $k^\text{m}$                    & $\nicefrac{0.1}{T^\text{m}}$         \\
        ~Sarcomere mean survival time  & $T^\text{m}$                    & $10\,\si{days}$                      \\
        Collagen:                                                                                               \\
        ~Homeostatic stretch           & $\lambda_\text{h}^\text{c}$     & $1.062$                              \\
        ~Growth gain                   & $k^\text{c}$                    & $\nicefrac{0.1}{T^\text{c}}$         \\
        ~Mean survival time            & $T^\text{c}$                    & $15\,\si{days}$                      \\
        \textit{Adaptive tolerance}                                                                             \\
        ~Model equation                & $\varepsilon_\mathcal{Q}$       & $10^{-9}$                            \\
        ~Error indication              & $\varepsilon_\mathcal{Q}$       & $10^{-5}$                            \\
        \textit{Prestressing}                                                                                   \\
        ~Prestress tolerance           & $\varepsilon_\text{pre}$        & $0.01\,\si{mm}$                      \\
        \noalign{\smallskip}\hline
    \end{tabular}
\end{table}

\subsection{Results}\label{sec:reslts_threed:results}

Table~\ref{tab:parameters} shows the material and simulation parameters used for the organ-scale example.
The mechanobiologically stable case ran for $1000\,\si{days}$ and the mechanobiologically unstable
case for $600\,\si{days}$. \revb{The baseline timestep size is $\Delta s = 0.5\,\si{days}$,
    which is a twentieth of the minimum involved turnover time constant.}

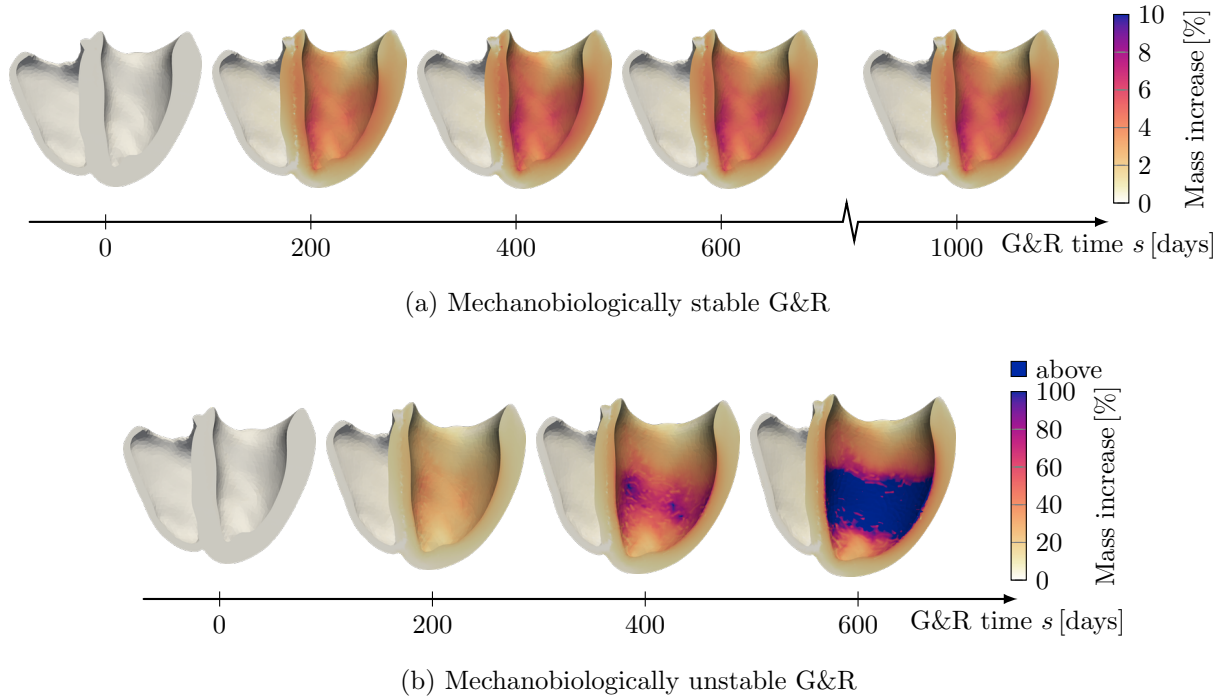
\begin{figure*}
    \import{figures/}{3d_result_plots.tex}
    \caption{Local mass increase at different times in the myocardium for
        (a) mechanobiologically stable and (b) unstable G\&R. \revb{The results stem from the
            full integration case with baseline timestep size. The differences to the other considered
            cases are negligibly small.}}
    \label{fig:3d_result_plots}
\end{figure*}

Figure~\ref{fig:3d_result_plots} shows the resulting G\&R after increasing the left ventricular pressure
for both loading scenarios. In the mechanobiological stable loading scenario, the maximum mass increase occurs at the endocardium of
the left ventricle and is around $6\,\si {\%}$, resulting in a configuration that is in mechanobiological
equilibrium. The mechanobiological unstable loading scenario does not result in a mechanobiological
equilibrated state within the considered timeframe. The maximum local mass increase is $130\,\si {\%}$
after $600\,\si{days}$ of unstable G\&R continuing to increase over time.

We ran the mechanobiologically stable and unstable loading cases for each adaptive strategy and
with a full history integration for comparison. As a reference solution, we additionally ran
the full integration case with a halved timestep size to compare the adaptive strategy errors
with the full integration. Halving the timestep size reduces the integration error
of Simpson's rule by a factor of $\frac{1}{32}$ \cite{Krommer1998a}, allowing to indicate the integration
error with a meaningful precision. A further reduction of the timestep size is infeasible for the 3D organ-scale example since
the computation time of the full integration case increases quadratically, and memory consumption increases linearly with the number of timesteps.

\begin{figure*}
    \import{figures/}{3d_gr_error_time_curves.tex}
    \caption{Relative errors of the myocyte fiber Cauchy stress, collagen mass fraction, and septum wall thickness
    for full history integration, model equation adaptive strategy with adaptive tolerance $\varepsilon_\mathcal{Q}=10^{-9}$
    and error indication adaptive strategy with adaptive tolerance $\varepsilon_\mathcal{Q}=10^{-5}$ for the 3D organ-scale
    example. The left ventricular pressure is increased at $s=0$ from baseline $120\,\si{mmHg}$ to (a) $140\,\si{mmHg}$ for
    mechanobiological stable and (b) $180\,\si{mmHg}$ for mechanobiological unstable G\&R.}
    \label{fig:gr_error_time_curves}
\end{figure*}
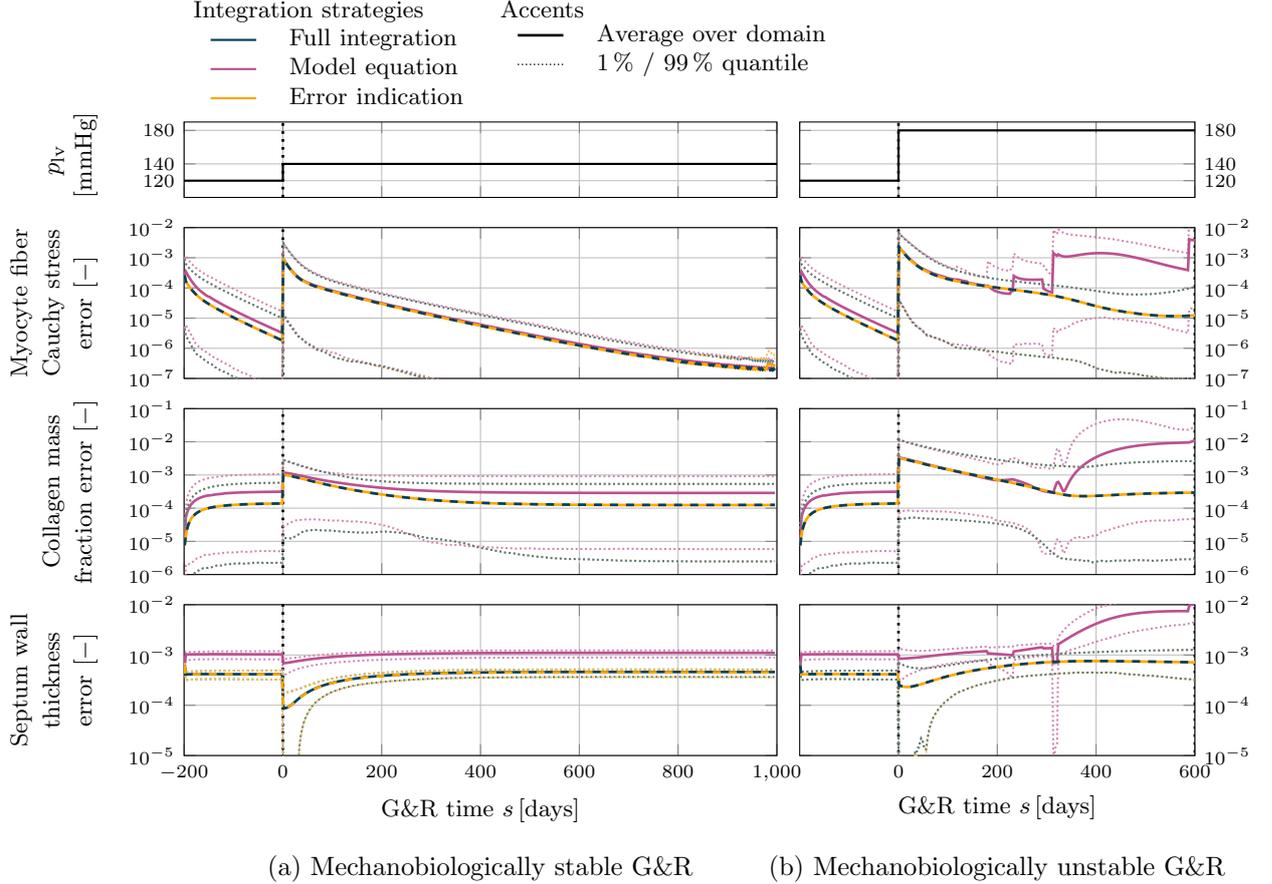

Figure~\ref{fig:gr_error_time_curves} shows the relative errors of the
myocyte fiber Cauchy stress, collagen mass fraction error, and the wall thickness of the septum of the considered cases. The
Cauchy stress and the mass fraction are evaluated at each integration
point. The wall thickness of the septum is computed for every
node on the endocardium of the septum and compared to the reference simulation. All those quantities
are shown as average over the domain, and spatial inhomogeneities
are indicated with the $1\,\%$ / $99\,\%$ quantiles of each quantity.

The model equation adaptive strategy can only partially reproduce the results of the full integration.
The error is especially pronounced in the late phase of mechanobiologically unstable G\&R dominated by the
adaptive integration.
In contrast, the error indication adaptive strategy can reproduce the results of the full integration for both loading scenarios.
Even at the end-stage of severe G\&R, the additional error of local and global quantities with
both adaptive strategies are small such that they can typically be neglected compared to other
errors in mechanobiological models.

\begin{figure*}
    \import{figures/}{3d_gr_time_curves.tex}
    \caption{History size and Newton step solution time of the 3D organ-scale example using
    full history integration, the
    model equation adaptive strategy \revb{with adaptive tolerance $\varepsilon_\mathcal{Q}=10^{-9}$}
    and error indication adaptive strategy \revb{with adaptive tolerance $\varepsilon_\mathcal{Q}=10^{-5}$
    (cf. Figure~\ref{fig:gr_error_time_curves})}. The pressure
    in the left ventricle is elevated from baseline $120\,\si{mmHg}$ to (a) $140\,\si{mmHg}$ (mechanobiologically
    stable G\&R) and (b) $180\,\si{mmHg}$ (mechanobiologically unstable G\&R).}
    \label{fig:gr_time_curves}
\end{figure*}

Figure~\ref{fig:gr_time_curves} shows the history size and the Newton step solution time.
The history size is the sum of the snapshots of all
constituents stored at each integration point.

The history size of the full integration increases \revc{linearly} since \revb{one snapshot is stored
    for each constituent and timestep.} At $s=1000\,\si{days}$\revb{, i.e., after 2400 timesteps,}
this results in a history size of $12\cdot10^3$ \revc{snapshots}
per quadrature point of the finite element discretization.
Like the history size, the Newton
step solution time also increases linearly.
The model equation adaptive theory is independent of the load, resulting in the same history size in
the myocardium for mechanobiologically stable and unstable G\&R. It plateaus at around $1000$ history snapshots with
the consequence that the computational costs also remain almost constant in time (cf., Newton step solution
time). When using the error indication adaptive strategy, the history size
depends on the local mechanobiological environment and is, therefore, locally inhomogeneous and different for mechanobiologically stable and unstable G\&R.
The final history size is also approximately $1000$ history snapshots with an almost constant Newton step
solution time.

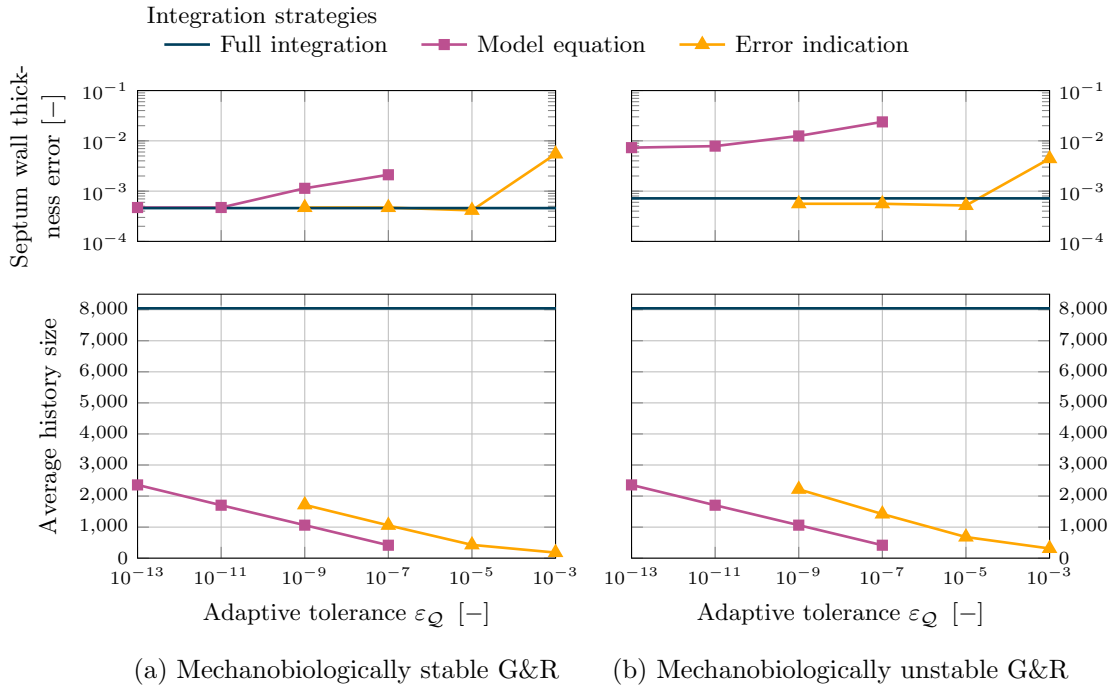
\begin{figure*}
    \import{figures/}{3d_error_per_tolerance.tex}
    \caption{Relative septum wall thickness error and history size for the organ-scale example using
        different adaptive tolerances at the time $s=600\,\si{days}$ for the (a) mechanobiologically
        stable and (b) mechanobiologically unstable loading scenario.}
    \label{fig:3d_error_per_tolerance}
\end{figure*}

Figure~\ref{fig:3d_error_per_tolerance} shows the relative septum wall thickness error and the
average history size at G\&R time $s=600\,\si{days}$ for different adaptive tolerances $\varepsilon_\mathcal{Q}$.
Note that the septum wall thickness is an organ-scale quantity and not
mechanobiologically regulated, i.e., the errors in this quantity accumulate over time.

The full history integration defines the lower bound of the relative septum wall thickness error.
The
error indication adaptive strategy reaches the lower bound with an adaptive tolerance of $10^{-5}$
for both loading scenarios. Reducing the adaptive tolerance does not further reduce the wall thickness
error since the base timestep size dominates the error.
It is just a coincidence that the error of the mechanobiological unstable G\&R case is slightly below
the lower bound.
The error of the model equation
adaptive strategy is higher than that of the error indication adaptive strategy. An adaptive tolerance
of $10^{-11}$ is needed for the mechanobiologically stable G\&R scenario to reduce the error to the lower bound.
For the mechanobiologically unstable G\&R scenario, the model equation adaptive strategy cannot
reduce the error to the lower bound within the considered tolerance range.

It is important to note that to decrease the integration error below the lower bound defined by the
full integration, it is necessary to decrease the base timestep size of the integration.

The average history sizes of both adaptive strategies in Figure~\ref{fig:3d_error_per_tolerance} are
roughly on a straight line on the
semilogarithmic plot with similar slope. Reducing the tolerance by a factor of $10^{-1}$ results
in a history size that is larger by around $300-400$ snapshots. The average history size is smaller
for the model equation adaptive strategy with the same adaptive tolerance compared to the error
indication adaptive strategy. In the latter, unstable G\&R also results in a larger history size than
for stable G\&R. Even for very small tolerances, the average history size is significantly smaller
compared to the full history integration case.

%% file: figures/3d_geometry_and_bc.tex
\begin{center}
    \begin{tikzpicture}
        \node[inner sep=0pt] (geometry_elementy) at (-2.5,0) {\includegraphics{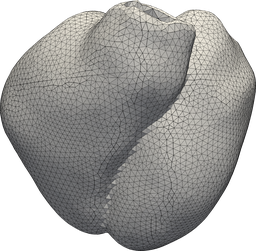}};
        \node[inner sep=0pt] (geometry_full) at (0,0) {\includegraphics{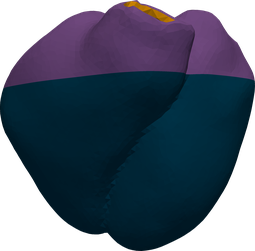}};
        \node[inner sep=0pt] (geometry_clipped) at (2.5,0) {\includegraphics{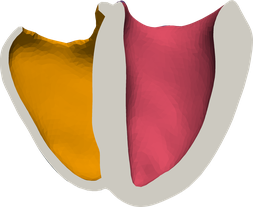}};

        \node[below] at (-2.5,-1.25) {Discretization};
        \node[below] at (0,-1.25) {Epicardium};
        \node[text width=3cm,align=center,below] at (2.5,-1.25) {Endocardium\\(clipped)};

        \begin{scope}[shift={(-0.5 cm,3 cm)}]
            \begin{scope}[shift={(-3 cm,0 cm)}]
                \coordinate (color1) at (0,0);
                \fill[color=dia4x1] ($(color1)+(-0.1,-0.1)$) rectangle++(0.2,0.2);
                \draw (color1)++(0.1,0) node[right] {Epicardium $\Gamma_{\text{epi}}$};

                \coordinate (color2) at (0,-0.4);
                \fill[color=dia4x2] ($(color2)+(-0.1,-0.1)$) rectangle++(0.2,0.2);
                \draw (color2)++(0.1,0) node[right] {Base $\Gamma_{\text{b}}$};
            \end{scope}
            \begin{scope}[shift={(0 cm,0 cm)}]
                \coordinate (color3) at (0,0.0);
                \fill[color=dia4x3] ($(color3)+(-0.1,-0.1)$) rectangle++(0.2,0.2);
                \draw (color3)++(0.1,0) node[right] {Left endocardium $\Gamma_{\text{lv}}$};

                \coordinate (color4) at (0,-0.4);
                \fill[color=dia4x4] ($(color4)+(-0.1,-0.1)$) rectangle++(0.2,0.2);
                \draw (color4)++(0.1,0) node[right] {Right endocardium $\Gamma_{\text{rv}}$};
            \end{scope}
        \end{scope}
    \end{tikzpicture}
\end{center}

%% file: figures/3d_result_plots.tex
\begin{center}
    \input{figures/images/cet_linear_worb_100_25_c53}
    \begin{tikzpicture}
        \begin{scope}[local bounding box=plot_scope_stable]
            \node[inner sep=0pt] (plot11) at (0,0) {\includegraphics{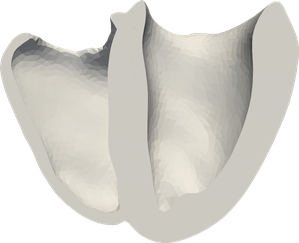}};
            \node[inner sep=0pt,right =0.13cm of plot11] (plot12) {\includegraphics{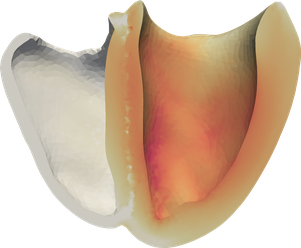}};
            \node[inner sep=0pt,right =0.13cm of plot12] (plot13) {\includegraphics{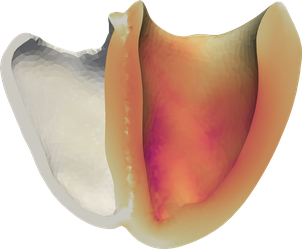}};
            \node[inner sep=0pt,right =0.13cm of plot13] (plot14) {\includegraphics{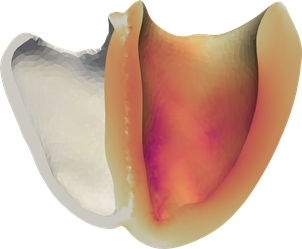}};
            \node[inner sep=0pt,right =0.6cm of plot14] (plot15) {\includegraphics{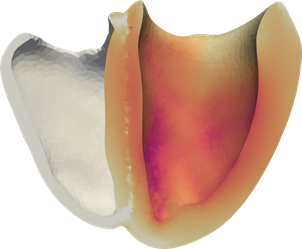}};

            \coordinate (start_axis) at (-1,-1.5);
            \draw[-latex,thick,line cap=round] (start_axis) -- ++(10.7,0) -- ++(0.05,0.25) -- ++(0.1,-0.5) -- ++(0.05,0.25) -- ++(3.3,0) node[below] {G\&R time $s\,[\si{days}]$};

            \draw (0,-1.5) ++(0,0.1) -- ++(0,-0.2) node[below] {$0$};
            \draw (2.7,-1.5) ++(0,0.1) -- ++(0,-0.2) node[below] {$200$};
            \draw (5.4,-1.5) ++(0,0.1) -- ++(0,-0.2) node[below] {$400$};
            \draw (8.1,-1.5) ++(0,0.1) -- ++(0,-0.2) node[below] {$600$};

            \draw (11.2,-1.5) ++(0,0.1) -- ++(0,-0.2) node[below] {$1000$};

            \begin{axis}[
                    scale only axis,
                    view={0}{90},
                    width=0.3cm,
                    height=2.5cm,
                    hide axis,
                    colormap/cetlinearworb,
                    colorbar,
                    colorbar style={%
                            anchor=north,
                            height=2.5cm,
                            width=0.2cm,
                            ytick={0,2,4,6,8,10},
                            ylabel=Mass increase\,$[\%]$
                        },
                    point meta min=0,
                    point meta max=10,
                    name=plotcolorbar,
                    at={($(plot15.east)+(0.5cm,0)$)},
                    anchor=west
                ]
            \end{axis}
        \end{scope}

        \begin{scope}[local bounding box=plot_scope_unstable, yshift=-5cm,xshift=1.5cm]

            \node[inner sep=0pt] (plot11) at (0,0) {\includegraphics{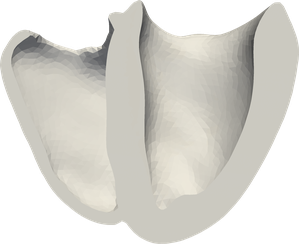}};
            \node[inner sep=0pt,right =0.13cm of plot11] (plot12) {\includegraphics{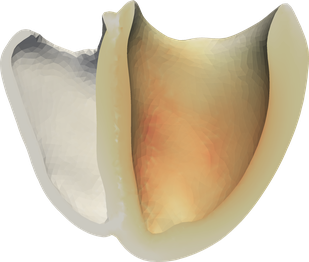}};
            \node[inner sep=0pt,right =0.13cm of plot12] (plot13) {\includegraphics{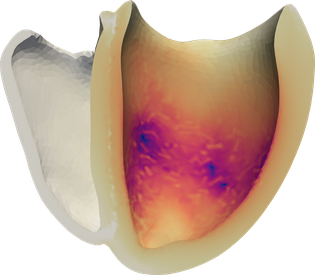}};
            \node[inner sep=0pt,right =0.13cm of plot13] (plot14) {\includegraphics{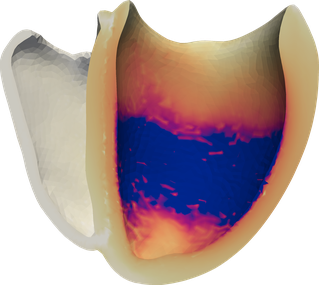}};

            \coordinate (start_axis) at (-1,-1.5);
            \draw[-latex,thick,line cap=round] (start_axis) -- ++(11.5,0) node[below] {G\&R time $s\,[\si{days}]$};

            \draw (0,-1.5) ++(0,0.1) -- ++(0,-0.2) node[below] {$0$};
            \draw (2.8,-1.5) ++(0,0.1) -- ++(0,-0.2) node[below] {$200$};
            \draw (5.6,-1.5) ++(0,0.1) -- ++(0,-0.2) node[below] {$400$};
            \draw (8.4,-1.5) ++(0,0.1) -- ++(0,-0.2) node[below] {$600$};

            \begin{axis}[
                    scale only axis,
                    view={0}{90},
                    width=0.3cm,
                    height=2.5cm,
                    hide axis,
                    colormap/cetlinearworb,
                    colorbar,
                    colorbar style={%
                            anchor=north,
                            height=2.5cm,
                            width=0.2cm,
                            ytick={0,20,40,60,80,100},
                            ylabel=Mass increase\,$[\%]$
                        },
                    point meta min=0,
                    point meta max=100,
                    name=plotcolorbar,
                    at={($(plot14.east)+(0.5cm,0)$)},
                    anchor=west
                ]
            \end{axis}

            \definecolor{maxcolorcetcolor}{RGB}{0,42,167}
            \draw[fill=maxcolorcetcolor] (plotcolorbar.north)++(0.075,0.2) -- ++(0.2cm,0) -- node[right] {above} ++(0,0.2cm) -- ++(-0.2cm,0) -- cycle;
        \end{scope}

        \draw ($(plot_scope_stable.south)+(0,-0.5)$) node{(a) Mechanobiologically stable G\&R};
        \draw ($(plot_scope_unstable.south)+(0,-0.5)$) node{(b) Mechanobiologically unstable G\&R};

    \end{tikzpicture}
\end{center}

%% file: figures/3d_gr_error_time_curves.tex
\begin{center}
    \begin{tikzpicture}[mydashed/.style={dashed,dash phase=3pt}]
        \begin{axis}[
                ticklabel style = {font=\footnotesize},
                scale only axis=true,
                width=7.8cm,
                height=1cm,
                xmin=-200, xmax=1000,
                ymin=100, ymax=190,
                ytick={120, 140, 180},
                xlabel=,
                ylabel=\small{\(p_\text{lv}\) \(\left[\si{mmHg}\right]\)},
                y label style={at={(axis description cs:-0.13,.5)},text width=1cm,align=center},
                xticklabels=\empty,
                name=lv_pressure_stable,
                anchor=north,
                grid=major,
            ]
            \addplot[color=black, line width=1pt, dotted] coordinates {(0,100) (0,190)};

            \addplot [thick, black] coordinates {
                    (-200,120)    (0,120)    (0,140)   (1000,140)
                };
        \end{axis}

        \begin{axis}[
                ticklabel style = {font=\footnotesize},
                scale only axis=true,
                width=7.8cm,
                height=2cm,
                xmin=-200, xmax=1000,
                ymin=1e-7, ymax=1e-2,
                ytick={1e-1,1e-2,1e-3,1e-4,1e-5,1e-6,1e-7,1e-8,1e-9},
                xlabel=,
                ylabel=\small{Myocyte fiber Cauchy stress error \(\left[-\right]\)},
                y label style={at={(axis description cs:-0.13,.5)},text width=2.5cm,align=center},
                xticklabels=\empty,
                name=myocyte_cauchy_stress_stable,
                at={($(lv_pressure_stable.south)+(0,-0.4cm)$)},
                anchor=north,
                grid=major,
                ymode=log,
            ]
            \addplot[color=black, line width=1pt, dotted] coordinates {(0,1e-9) (0,1e-1)};

            \plotquantiles{figures/data/3d/gr1_p140_model_equation_subset0_fiber0_cauchy_stress_error.csv}{dia2-2}{time}
            \plotquantiles{figures/data/3d/gr1_p140_higher_order_subset0_fiber0_cauchy_stress_error.csv}{dia3-2}{time}
            \plotquantiles[dashed]{figures/data/3d/gr1_p140_none_subset0_fiber0_cauchy_stress_error.csv}{dia1-2}{time}
        \end{axis}

        \begin{axis}[
                ticklabel style = {font=\footnotesize},
                scale only axis=true,
                width=7.8cm,
                height=2.2cm,
                xmin=-200, xmax=1000,
                ymin=1e-6, ymax=1e-1,
                ytick={1e-1,1e-2,1e-3,1e-4,1e-5,1e-6,1e-7},
                xlabel=,
                ylabel=\small{Collagen mass fraction error \(\left[-\right]\)},
                y label style={at={(axis description cs:-0.13,.5)},text width=2.5cm,align=center},
                xticklabels=\empty,
                name=collagen_stable,
                at={($(myocyte_cauchy_stress_stable.south)+(0,-0.4cm)$)},
                ymode=log,
                grid=major,
                anchor=north,
            ]
            \addplot[color=black, line width=1pt, dotted] coordinates {(0,1e-7) (0,1e-1)};

            \plotquantiles{figures/data/3d/gr1_p140_model_equation_subset0_collagen_mass_fraction.csv}{dia2-2}{time}
            \plotquantiles{figures/data/3d/gr1_p140_higher_order_subset0_collagen_mass_fraction.csv}{dia3-2}{time}
            \plotquantiles[dashed]{figures/data/3d/gr1_p140_none_subset0_collagen_mass_fraction.csv}{dia1-2}{time}
        \end{axis}

        \begin{axis}[
                ticklabel style = {font=\footnotesize},
                scale only axis=true,
                width=7.8cm,
                height=2cm,
                name=cell_mms_uncoupled_left,
                xmin=-200, xmax=1000,
                ymin=1e-5, ymax=1e-2,
                ytick={1e-2,1e-3,1e-4,1e-5,1e-6,1e-7},
                xtick={-200,0,200,400,600,800,1000},
                xlabel=\small{G\&R time \(s \left[\si{days}\right]\)},
                ylabel=\small{Septum wall thickness error \(\left[-\right]\)},
                y label style={at={(axis description cs:-0.13,.5)},text width=2.5cm,align=center},
                name=wall_thickness_stable,
                at={($(collagen_stable.south)+(0,-0.4cm)$)},
                anchor=north,
                grid=major,
                ymode=log,
            ]
            \addplot[color=black, line width=1pt, dotted] coordinates {(0,1e-9) (0,1e-1)};

            \plotquantiles{figures/data/3d/gr1_p140_none_wall_thickness_septum_error.csv}{dia1-2}{time}
            \plotquantiles{figures/data/3d/gr1_p140_model_equation_wall_thickness_septum_error.csv}{dia2-2}{time}
            \plotquantiles[dashed]{figures/data/3d/gr1_p140_higher_order_wall_thickness_septum_error.csv}{dia3-2}{time}
        \end{axis}

        \begin{axis}[
                ticklabel style = {font=\footnotesize},
                scale only axis=true,
                width=5.2cm,
                height=1cm,
                xmin=-200, xmax=600,
                ymin=100, ymax=190,
                ytick={120, 140, 180},
                xlabel=,
                ylabel=,
                y label style={at={(axis description cs:-0.13,.5)},text width=1cm,align=center},
                xticklabels=\empty,
                name=lv_pressure_unstable,
                at={($(lv_pressure_stable.east)+(0.3cm,0cm)$)},
                anchor=west,
                grid=major,
                yticklabel pos=right,
            ]
            \addplot[color=black, line width=1pt, dotted] coordinates {(0,100) (0,190)};
            \addplot [thick, black] coordinates {
                    (-200,120)    (0,120)    (0,180)   (600,180)
                };
        \end{axis}

        \begin{axis}[
                ticklabel style = {font=\footnotesize},
                scale only axis=true,
                width=5.2cm,
                height=2cm,
                xmin=-200, xmax=600,
                ymin=1e-7, ymax=1e-2,
                ytick={1e-1,1e-2,1e-3,1e-4,1e-5,1e-6,1e-7,1e-8,1e-9},
                xlabel=,
                yticklabel pos=right,
                ylabel=,
                y label style={at={(axis description cs:-0.13,.5)},text width=3.5cm,align=center},
                xticklabels=\empty,
                anchor=north,
                grid=major,
                ymode=log,
                name=myocyte_cauchy_stress_unstable,
                at={($(myocyte_cauchy_stress_stable.east)+(0.3cm,0cm)$)},
                anchor=west,
            ]
            \addplot[color=black, line width=1pt, dotted] coordinates {(0,1e-9) (0,1e-1)};
            \addplot[color=black, line width=1pt, dotted] coordinates {(600,1e-9) (600,1e-1)};

            \plotquantiles[restrict x to domain=-200:601]{figures/data/3d/gr1_p180_model_equation_subset0_fiber0_cauchy_stress_error.csv}{dia2-2}{time}
            \plotquantiles[restrict x to domain=-200:601]{figures/data/3d/gr1_p180_higher_order_subset0_fiber0_cauchy_stress_error.csv}{dia3-2}{time}
            \plotquantiles[restrict x to domain=-200:601, dashed]{figures/data/3d/gr1_p180_none_subset0_fiber0_cauchy_stress_error.csv}{dia1-2}{time}
        \end{axis}

        \begin{axis}[
                ticklabel style = {font=\footnotesize},
                scale only axis=true,
                width=5.2cm,
                height=2.2cm,
                xmin=-200, xmax=600,
                ymin=1e-6, ymax=1e-1,
                ytick={1e-1,1e-2,1e-3,1e-4,1e-5,1e-6,1e-7},
                xlabel=,
                yticklabel pos=right,
                ylabel=,
                y label style={at={(axis description cs:-0.13,.5)},text width=3.5cm,align=center},
                xticklabels=\empty,
                name=collagen_unstable,
                at={($(myocyte_cauchy_stress_unstable.south)+(0,-0.4cm)$)},
                ymode=log,
                grid=major,
                anchor=north,
            ]
            \addplot[color=black, line width=1pt, dotted] coordinates {(0,1e-7) (0,1e-1)};
            \addplot[color=black, line width=1pt, dotted] coordinates {(600,1e-7) (600,1e-1)};
            \plotquantiles[restrict x to domain=-200:601]{figures/data/3d/gr1_p180_model_equation_subset0_collagen_mass_fraction.csv}{dia2-2}{time}
            \plotquantiles[restrict x to domain=-200:601]{figures/data/3d/gr1_p180_higher_order_subset0_collagen_mass_fraction.csv}{dia3-2}{time}
            \plotquantiles[restrict x to domain=-200:601, dashed]{figures/data/3d/gr1_p180_none_subset0_collagen_mass_fraction.csv}{dia1-2}{time}
        \end{axis}

        \begin{axis}[
                ticklabel style = {font=\footnotesize},
                scale only axis=true,
                width=5.2cm,
                height=2cm,
                name=cell_mms_uncoupled_left,
                xmin=-200, xmax=600,
                ymin=1e-5, ymax=1e-2,
                ytick={1e-2,1e-3,1e-4,1e-5,1e-6,1e-7},
                xlabel=\small{G\&R time \(s \left[\si{days}\right]\)},
                yticklabel pos=right,
                xtick={0,200,400,600,800,1000},
                ylabel=,
                y label style={at={(axis description cs:-0.13,.5)},text width=3.5cm,align=center},
                name=wall_thickness_unstable,
                at={($(collagen_unstable.south)+(0,-0.4cm)$)},
                anchor=north,
                grid=major,
                ymode=log,
            ]
            \addplot[color=black, line width=1pt, dotted] coordinates {(0,1e-7) (0,1e-1)};
            \addplot[color=black, line width=1pt, dotted] coordinates {(600,1e-7) (600,1e-1)};
            \plotquantiles[restrict x to domain=-200:601]{figures/data/3d/gr1_p180_model_equation_wall_thickness_septum_error.csv}{dia2-2}{time}
            \plotquantiles[restrict x to domain=-200:601]{figures/data/3d/gr1_p180_higher_order_wall_thickness_septum_error.csv}{dia3-2}{time}
            \plotquantiles[restrict x to domain=-200:601, dashed]{figures/data/3d/gr1_p180_none_wall_thickness_septum_error.csv}{dia1-2}{time}
        \end{axis}

        \node[anchor=south west,align=left,text width=\textwidth-2cm] at (lv_pressure_stable.north west) {%
            {\small\begin{tabular}[t]{lll}
                        Integration strategies     & Accents & ~ \\
                        \begin{tabular}[t]{cl}
                            \begin{tikzpicture}
                                \fill[color=white] (0,-0.075) rectangle(0.6,0.15);
                                \draw[thick, dia1-2] (0,0) -- ++(0.6,0);
                            \end{tikzpicture}  & Full integration \\
                            \begin{tikzpicture}
                                \fill[color=white] (0,-0.075) rectangle(0.6,0.15);
                                \draw[thick, dia2-2] (0,0) -- ++(0.6,0);
                            \end{tikzpicture} & Model equation   \\
                            \begin{tikzpicture}
                                \fill[color=white] (0,-0.075) rectangle(0.6,0.15);
                                \draw[thick, dia3-2] (0,0) -- ++(0.6,0);
                            \end{tikzpicture} & Error indication \\
                        \end{tabular} &
                        \begin{tabular}[t]{cl}
                            \begin{tikzpicture}
                                \fill[color=white] (0,-0.075) rectangle(0.6,0.15);
                                \draw[line width=1pt,black] (0,0.0) -- ++(0.6,0);
                            \end{tikzpicture} & Average over domain         \\
                            \begin{tikzpicture}
                                \fill[color=white] (0,-0.075) rectangle(0.6,0.15);
                                \draw[line width=0.5pt,black, densely dotted] (0,0.0) -- ++(0.6,0);
                            \end{tikzpicture} & $1\,\%$ / $99\,\%$ quantile \\
                        \end{tabular} &
                        \begin{tabular}[t]{cl}
                            \begin{tikzpicture}
                                \draw[line width=1pt, densely dotted,black] (0.1,-0.2) -- ++(0,0.25);
                            \end{tikzpicture} & Start of hypertension
                        \end{tabular}
                    \end{tabular}}
        };

        \draw ($(wall_thickness_stable.south)+(0,-1.5)$) node{(a) Mechanobiologically stable G\&R};
        \draw ($(wall_thickness_unstable.south)+(0,-1.5)$) node{(b) Mechanobiologically unstable G\&R};
    \end{tikzpicture}
\end{center}

%% file: figures/3d_gr_time_curves.tex
\begin{center}
    \begin{tikzpicture}[mydashed/.style={dashed,dash phase=3pt}]
        \begin{axis}[
                ticklabel style = {font=\footnotesize},
                scale only axis=true,
                width=7.8cm,
                height=3.5cm,
                name=cell_mms_uncoupled_left,
                scaled y ticks=base 10:-3,
                xmin=-190, xmax=1000,
                ymin=0, ymax=12000,
                xlabel=,
                ylabel=\small{History size \(\left[\si{-}\right]\)},
                ytick={0,2000,4000,6000,8000,10000,12000},
                xticklabels=\empty,
                name=history_size_stable,
                grid=major,
                anchor=north,
            ]
            \addplot[color=black, line width=1pt, dotted] coordinates {(0,0) (0,12000)};

            \addplot[mark=none, line width=1pt, color=dia2-2]
            table [x=time, y=average, col sep=comma]
                {figures/data/3d/gr1_p140_model_equation_subset0_history_size.csv};

            \plotquantiles{figures/data/3d/gr1_p140_higher_order_subset0_history_size.csv}{dia3-2}{time}

            \addplot[mark=none, line width=1pt, color=dia1-2]
            table [x=time, y=average, col sep=comma]
                {figures/data/3d/gr1_p140_none_subset0_history_size.csv};
        \end{axis}

        \begin{axis}[
                ticklabel style = {font=\footnotesize},
                scale only axis=true,
                width=7.8cm,
                height=3.5cm,
                xmin=-190, xmax=1000,
                ymin=0, ymax=22,
                ytick={0,5,10,15,20},
                ylabel=\small{Newton step solution time \(\left[\si{s}\right]\)},
                y label style={at={(axis description cs:-0.07,.5)},text width=3cm,align=center},
                xlabel=\small{G\&R time \(s \left[\si{days}\right]\)},
                name=newton_step_solution_time_stable_top,
                grid=major,
                at={($(history_size_stable.south)+(0,-0.4cm)$)},
                anchor=north,
            ]
            \addplot[color=black, line width=1pt, dotted] coordinates {(0,0) (0,22)};
            \addplot[mark=none, line width=1pt, color=dia2-2] table [x=time, y=newton_solution_time, col sep=comma]
                {figures/data/3d/gr1_p140_model_equation_timestep_runtime.csv};
            \addplot[mark=none, line width=1pt, color=dia3-2] table [x=time, y=newton_solution_time, col sep=comma]
                {figures/data/3d/gr1_p140_higher_order_timestep_runtime.csv};

            \addplot[mark=none, line width=1pt, color=dia1-2 ] table [x=time, y=newton_solution_time, col sep=comma]
                {figures/data/3d/gr1_p140_none_timestep_runtime.csv};
        \end{axis}

        \begin{axis}[
                ticklabel style = {font=\footnotesize},
                scale only axis=true,
                width=5.2cm,
                height=3.5cm,
                xmin=-190, xmax=600,
                ymin=0, ymax=12000,
                scaled y ticks=base 10:-3,
                xlabel=,
                ylabel=,
                yticklabel pos=right,
                ytick={0,2000,4000,6000,8000,10000,12000},
                xticklabels=\empty,
                name=history_size_unstable,
                grid=major,
                at={($(history_size_stable.east)+(0.3cm,0)$)},
                anchor=west,
            ]
            \addplot[color=black, line width=1pt, dotted] coordinates {(0,0) (0,12000)};
            \addplot[mark=none, line width=1pt, color=dia2-2]
            table [x=time, y=average, col sep=comma]
                {figures/data/3d/gr1_p180_model_equation_subset0_history_size.csv};

            \plotquantiles[restrict x to domain=-200:600]{figures/data/3d/gr1_p180_higher_order_subset0_history_size.csv}{dia3-2}{time}

            \addplot[mark=none, line width=1pt, color=dia1-2]
            table [x=time, y=average, col sep=comma]
                {figures/data/3d/gr1_p180_none_subset0_history_size.csv};
        \end{axis}

        \begin{axis}[
                ticklabel style = {font=\footnotesize},
                scale only axis=true,
                width=5.2cm,
                height=3.5cm,
                xmin=-190, xmax=600,
                ymin=0, ymax=22,
                ytick={0,5,10,15,20},
                yticklabel pos=right,
                xlabel=\small{G\&R time \(s \left[\si{days}\right]\)},
                ylabel=,
                y label style={at={(axis description cs:-0.1,.5)},text width=3cm,align=center},
                name=newton_step_solution_time_unstable_top,
                grid=major,
                at={($(history_size_unstable.south)+(0,-0.4cm)$)},
                anchor=north,
            ]
            \addplot[color=black, line width=1pt, dotted] coordinates {(0,0) (0,22)};

            \addplot[mark=none, line width=1pt, color=dia2-2,restrict x to domain=-200:600] table [x=time, y=newton_solution_time, col sep=comma]
                {figures/data/3d/gr1_p180_model_equation_timestep_runtime.csv};

            \addplot[mark=none, line width=1pt, color=dia3-2,restrict x to domain=-200:600 ] table [x=time, y=newton_solution_time, col sep=comma]
                {figures/data/3d/gr1_p180_higher_order_timestep_runtime.csv};

            \addplot[mark=none, line width=1pt, color=dia1-2,restrict x to domain=-200:600] table [x=time, y=newton_solution_time, col sep=comma]
                {figures/data/3d/gr1_p180_none_timestep_runtime.csv};
        \end{axis}

        \node[anchor=south west,align=left,text width=\textwidth-2cm] at ($(history_size_stable.north west)+(0,0.1)$) {%
            {\small\begin{tabular}[t]{lll}
                        Integration strategies     & Accents & ~ \\
                        \begin{tabular}[t]{cl}
                            \begin{tikzpicture}
                                \fill[color=white] (0,-0.075) rectangle(0.6,0.15);
                                \draw[thick, dia1-2] (0,0) -- ++(0.6,0);
                            \end{tikzpicture} & Full integration \\
                            \begin{tikzpicture}
                                \fill[color=white] (0,-0.075) rectangle(0.6,0.15);
                                \draw[thick, dia2-2] (0,0) -- ++(0.6,0);
                            \end{tikzpicture} & Model equation   \\
                            \begin{tikzpicture}
                                \fill[color=white] (0,-0.075) rectangle(0.6,0.15);
                                \draw[thick, dia3-2] (0,0) -- ++(0.6,0);
                            \end{tikzpicture} & Error indication \\
                        \end{tabular} &
                        \begin{tabular}[t]{cl}
                            \begin{tikzpicture}
                                \fill[color=white] (0,-0.075) rectangle(0.6,0.15);
                                \draw[line width=1pt,black] (0,0.0) -- ++(0.6,0);
                            \end{tikzpicture} & Average over domain         \\
                            \begin{tikzpicture}
                                \fill[color=white] (0,-0.075) rectangle(0.6,0.15);
                                \draw[line width=0.5pt,black, densely dotted] (0,0.0) -- ++(0.6,0);
                            \end{tikzpicture} & $1\,\%$ / $99\,\%$ quantile \\
                        \end{tabular} &
                        \begin{tabular}[t]{cl}
                            \begin{tikzpicture}
                                \draw[line width=1pt, densely dotted,black] (0.1,-0.2) -- ++(0,0.25);
                            \end{tikzpicture} & Start of hypertension
                        \end{tabular}
                    \end{tabular}}
        };

        \draw ($(newton_step_solution_time_stable_top.south)+(0,-1.5)$) node{(a) Mechanobiologically stable G\&R};
        \draw ($(newton_step_solution_time_unstable_top.south)+(0,-1.5)$) node{(b) Mechanobiologically unstable G\&R};
    \end{tikzpicture}
\end{center}

%% file: figures/3d_error_per_tolerance.tex
\begin{center}
    \begin{tikzpicture}
        \begin{axis}[
                ticklabel style = {font=\footnotesize},
                scale only axis=true,
                width=5.5cm,
                height=2cm,
                ylabel=\small{Septum wall thickness error $\left[\si{-}\right]$},
                xlabel=,
                xticklabels=\empty,
                y label style={at={(axis description cs:-0.15,.5)},text width=3.5cm,align=center},
                xmin=1e-13, xmax=1e-3,
                ymin=1e-4, ymax=1e-1,
                xtick={1e-3,1e-5,1e-7,1e-9,1e-11,1e-13},
                name=septum_wall_thickness_error_tolerance_stable,
                grid=major,
                xmode=log,
                ymode=log,
            ]
            \addplot[mark=square*, line width=1pt, color=dia2-2,mark options={scale=0.8} ] plot table [x=tolerance, y=average, col sep=comma]
                {figures/data/3d/model_equation_p140_wall_thickness_septum_error_per_tolerance.csv};
            \label{plot:model_equation}

            \addplot[mark=triangle*, line width=1pt, color=dia3-2,mark options={scale=1.2} ] plot table [x=tolerance, y=average, col sep=comma]
                {figures/data/3d/higher_order_p140_wall_thickness_septum_error_per_tolerance.csv};
            \label{plot:error_estimation}

            \addplot[mark=none, line width=1pt, color=dia1-2] plot coordinates
                {(1e-13, 4.596231273334623916e-04) (1, 4.596231273334623916e-04)};
            \label{plot:full_integration}

        \end{axis}

        \begin{axis}[
                ticklabel style = {font=\footnotesize},
                scale only axis=true,
                width=5.5cm,
                height=3.5cm,
                xlabel=\small{Adaptive tolerance $\varepsilon_\mathcal{Q}~\left[\si{-}\right]$},
                ylabel=\small{Average history size},
                y label style={text width=3.5cm,align=center},
                xmin=1e-13, xmax=1e-3,
                ymin=0, ymax=8500,
                xtick={1e-3,1e-5,1e-7,1e-9,1e-11,1e-13},
                ytick={0,1000,2000,3000,4000,5000,6000,7000,8000},
                name=history_size_tolerance_stable,
                at={($(septum_wall_thickness_error_tolerance_stable.south)+(0,-0.7cm)$)},
                anchor=north,
                grid=major,
                xmode=log,
            ]

            \addplot[mark=none, line width=1pt, color=dia1-2 ] plot coordinates
                {(1e-13, 8.045000000000000000e+03) (1, 8.045000000000000000e+03)};

            \addplot[mark=square*, line width=1pt, color=dia2-2,mark options={scale=0.8}] plot table [x=tolerance, y=average, col sep=comma]
                {figures/data/3d/model_equation_p140_subset0_history_size_per_tolerance.csv};

            \addplot[mark=triangle*, line width=1pt, color=dia3-2,mark options={scale=1.2} ] plot table [x=tolerance, y=average, col sep=comma]
                {figures/data/3d/higher_order_p140_subset0_history_size_per_tolerance.csv};
        \end{axis}

        \begin{axis}[
                ticklabel style = {font=\footnotesize},
                scale only axis=true,
                width=5.5cm,
                height=2cm,
                ylabel=,
                xlabel=,
                xticklabels=\empty,
                y label style={at={(axis description cs:-0.15,.5)},text width=3.5cm,align=center},
                xmin=1e-13, xmax=1e-3,
                ymin=1e-4, ymax=1e-1,
                xtick={1e-3,1e-5,1e-7,1e-9,1e-11,1e-13},
                name=septum_wall_thickness_error_tolerance_unstable,
                at={($(septum_wall_thickness_error_tolerance_stable.east)+(1cm,0)$)},
                anchor=west,
                grid=major,
                xmode=log,
                ymode=log,
                yticklabel pos=right,
            ]
            \addplot[mark=square*, line width=1pt, color=dia2-2,mark options={scale=0.8} ] plot table [x=tolerance, y=average, col sep=comma]
                {figures/data/3d/model_equation_p180_wall_thickness_septum_error_per_tolerance.csv};

            \addplot[mark=triangle*, line width=1pt, color=dia3-2,mark options={scale=1.2} ] plot table [x=tolerance, y=average, col sep=comma]
                {figures/data/3d/higher_order_p180_wall_thickness_septum_error_per_tolerance.csv};

            \addplot[mark=none, line width=1pt, color=dia1-2 ] plot coordinates
                {(1e-13, 7.178583864203768542e-04) (1, 7.178583864203768542e-04)};

        \end{axis}

        \begin{axis}[
                ticklabel style = {font=\footnotesize},
                scale only axis=true,
                width=5.5cm,
                height=3.5cm,
                xlabel=\small{Adaptive tolerance $\varepsilon_\mathcal{Q}~\left[\si{-}\right]$},
                ylabel=,
                y label style={at={(axis description cs:-0.18,.5)},text width=3.5cm,align=center},
                xmin=1e-13, xmax=1e-3,
                ymin=0, ymax=8500,
                xtick={1e-3,1e-5,1e-7,1e-9,1e-11,1e-13},
                ytick={0,1000,2000,3000,4000,5000,6000,7000,8000},
                name=history_size_tolerance_unstable,
                at={($(septum_wall_thickness_error_tolerance_unstable.south)+(0,-0.7cm)$)},
                anchor=north,
                grid=major,
                xmode=log,
                yticklabel pos=right,
            ]

            \addplot[mark=none, line width=1pt, color=dia1-2 ] plot coordinates
                {(1e-13, 8.045000000000000000e+03) (1, 8.045000000000000000e+03)};
            \addplot[mark=square*, line width=1pt, color=dia2-2,mark options={scale=0.8} ] plot table [x=tolerance, y=average, col sep=comma]
                {figures/data/3d/model_equation_p180_subset0_history_size_per_tolerance.csv};

            \addplot[mark=triangle*, line width=1pt, color=dia3-2,mark options={scale=1.2} ] plot table [x=tolerance, y=average, col sep=comma]
                {figures/data/3d/higher_order_p180_subset0_history_size_per_tolerance.csv};
        \end{axis}

        \node[anchor=south west,align=left,text width=\textwidth-2cm] at ($(septum_wall_thickness_error_tolerance_stable.north west)$) {%
            \begin{minipage}[t]{0.3\textwidth}
                \small{Integration strategies}
                {\small
                    \begin{tabular}{lll}
                        \ref{plot:full_integration} Full integration &
                        \ref{plot:model_equation} Model equation     &
                        \ref{plot:error_estimation} Error indication
                    \end{tabular}
                }
            \end{minipage}
        };

        \draw ($(history_size_tolerance_stable.south)+(0,-1.5)$) node{(a) Mechanobiologically stable G\&R};
        \draw ($(history_size_tolerance_unstable.south)+(0,-1.5)$) node{(b) Mechanobiologically unstable G\&R};
    \end{tikzpicture}
\end{center}

%% file: sections/discussion.tex
\section{Discussion}\label{sec:discussion}

We presented two strategies to adaptively integrate the history variables in constrained mixture models
with the goal of drastically reducing the computational effort of such models. The strategies exploit the
fact that the influence of tissue once deposited decreases over time through degradation. We analyzed
the strategies on a tissue patch in reduced dimensions and demonstrated that both strategies
enable the simulation of long-running G\&R models while keeping the error in both local and global
quantities below physiological relevant scales on a level comparable to the full integration.

\subsection{Strategy A: Model equation}

We presented a strategy that adapts the history integration by approximating the error with
analytical solution during tissue maintenance.
The strategy does not take into account the additional mass deposition if the tissue is not in homeostasis.

This method should be the method of choice if it is particularly important to predict the computational
costs of the model evaluation in advance. The history size only depends on the timestep size, the
degradation time constant, and the adaptive tolerance and can, therefore, typically be computed
ahead of the simulation. This comes with the downside that the additional integration error
is generally larger than the prescribed adaptive tolerance for severe G\&R scenarios.

\subsection{Strategy B: Error indication}

We also presented a strategy that uses a higher-order numerical integration scheme to compute the
error. This strategy uses the actual integrals that need to be solved in constrained mixture models.
As a result, the history size is spatially inhomogeneous, i.e., parts with higher mass deposition
store more history snapshots than those with less dominant mass deposition. The available computing
resources can, therefore, be optimally used.
The adaptive tolerance directly controls the additional error in local G\&R quantities and as,
a consequence, also organ-scale quantities. This makes this adaptive strategy particularly useful
for severe G\&R.

A downside of the adaptive strategy is that it is hardly possible to compute the
history size ahead of the simulation time. It is, therefore, difficult to estimate the needed
computational resources (e.g., the amount of needed peak memory).

\subsection{Computational costs}\label{sec:discussion:comp_costs}
We ran each simulation on one node of our in-house Linux cluster with AMD Epyc 9354 Zen 4 with $384\,\si{GB}$ RAM and
$32$ cores at $3.25\,\si{GHz}$. The non-adaptive history integration needed for mechanobiological stable
G\&R $253\,\si{GB}$ RAM, compared to $39\,\si{GB}$ for error indication and for model equation adaptive
strategy. The total simulation time for the $1000\,\si{days}$ period of stable G\&R was $31\,\si{h}$ for
the non-adaptive case compared to around $10\,\si{h}$ for the model equation and error indication
adaptive strategy.

Both simulation time and memory consumption correlate with the number of snapshots in the history.
With our adaptive strategies, the history size plateaus and remains fairly constant afterward\revb{, i.e.,
    the computational complexity of the memory requirements is $\mathcal{O}(1)$ with the number of timesteps. As a consequence,
    the solution time of one timestep is also constant in time. Hence, solving $n$ timesteps has a
    computation time complexity of $\mathcal{O}(n)$ as typically expected from a transient problem.}
For the non-adaptive
simulation, the memory consumption and the \revc{timestep} solution time grow linearly \revc{with the number of timesteps},
resulting in a computation time complexity
of $\mathcal{O}(\revc{n^2})$, hence \revc{eventually} reaching the boundary of \revc{any} computing hardware
\revb{in terms of memory consumption and evaluation time.
    To compare the simulation results, we limited our organ-scale example
    to be small enough to evaluate it with the non-adaptive integration. Larger meshes or longer
    G\&R periods, e.g., needed for studying mechanobiological stability or reversal,
    quickly make the non-adaptive integration infeasible. }

With our choice of tolerances, the increase of the evaluation time in the adaptive cases remains smaller
than the solution time of the iterative linear solver. Hence, the model solution time is dominated
by the linear solver. The evaluation of the whole model remains, therefore, comparable to models
that just integrate local ordinary differential equations, like homogenized constrained mixture
models \citep{Cyron2016a} and kinematic growth models \citep{Rodriguez1994a}.

\subsection{Limitations}\label{sec:discussion:limitations}
The topic of this paper was the introduction and investigation of two approaches to control the
\emph{additional} error of the adaptive integration. One needs to keep in mind that this is
not the total integration error. The latter is controlled by the timestep,
which should be small enough to ensure that the integration error is below
physiological relevant scales. The adaptive tolerance should be chosen small enough to ensure that the
\emph{additional} error of the adaptive integration is negligible in organ-scale G\&R results. This
is particularly important since ogan-scale quantities like the wall thickness, lumen, or organ mass
are not directly mechanobiologically controlled, and errors made during model evaluation accumulate
over time.

For simplicity, we assumed a simple Poisson degradation process for the tissue constituents. However,
processes leading to cell apoptosis and tissue degradation are generally complex and can depend on
the current mechanobiological environment. Adaptive integration of the history variables can still
result in a significant reduction of computational costs, especially when using the error indication
adaptive strategy, which naturally can handle any degradation rules.

In our implementation, we assume that constituents that do turnover are fibers that only contribute to
the mixture stress in their preferred direction and that new fibers are deposited in the same
direction \citep{Braeu2016a,Mousavi2017a,Braeu2019a,Mousavi2019a,Gebauer2023a}. This is not a limitation of
the proposed adaptive integration strategies. We expect similar speedup results for \reva{an
    implementation of a general 3D constituent with turnover following equation (\ref{eqn:current_constituent_stress_response}).
    That allows modeling fiber reorientation due to G\&R and also enables fiber dispersion models \cite{Gasser2006a}
    for predicting increased dispersion during disease progression \cite{Eriksson2013a}.}

In our model, the driving factor of G\&R is the difference in the Cauchy-stress of the constituent.
However, other G\&R stimuli like wall shear stress in blood vessels \cite{Rosen1974a,Humphrey2008a} or infiltration of
inflammatory cells \cite{Intengan2001a,Humphrey2008a,Latorre2018b} also play an important role in soft tissue G\&R. Our adaptive strategies
should also capture these cases, especially the error indication adaptive strategy where
additional stimuli are directly considered during adaption.

%% file: sections/appendix.tex
\section{Newton-Cotes integration rules}
\label{app:newton_cotes}

The global time-stepping of the finite element framework often results in equidistant timesteps. To
integrate the history variables in constrained mixture models, closed Newton-Cotes integration rules,
i.e., integration rules with equidistant quadrature points need to be used. We denote the
Newton-Cotes integration of the function $\bs{\mathcal{F}}$ with $n$ points on the interval
$\mathcal{I}=[a,b]$ with $\mathcal{Q}_n^\mathcal{I} (\bs{\mathcal{F}})$. The equidistant
quadrature points are at
\begin{align*}
    s_i = a + (i-1) \frac{b-a}{n-1}, \quad i=1,2,\dots,n
\end{align*}
and the function $\bs{\mathcal{F}}$ evaluated at $s_i$ is denoted as $\bs{\mathcal{F}}_i$.

The trapezoidal rule \cite{Krommer1998a} uses just the endpoints of the interval that we use in the first timestep
where only two snapshots are available. The quadrature is given as
\begin{align*}
    \mathcal{Q}_2^\mathcal{I} (\bs{\mathcal{F}}) & = \frac{b-a}{2} ( \bs{\mathcal{F}}_1 + \bs{\mathcal{F}}_2 ).
\end{align*}

Typically, during integration, we apply the composite Simpson's rule \cite{Krommer1998a} with an additional midpoint, i.e.,
\begin{align*}
    \mathcal{Q}_3^\mathcal{I} (\bs{\mathcal{F}}) & = \frac{b-a}{6} ( \bs{\mathcal{F}}_1 + 4\bs{\mathcal{F}}_2  + \bs{\mathcal{F}}_3 ).
\end{align*}

The Boole's integration \cite{Krommer1998a} is used to approximate the integration error to adaptively integrate the
G\&R history. The quadrature rule is
\begin{align*}
    \mathcal{Q}_5^\mathcal{I} (\bs{\mathcal{F}}) & = \frac{b-a}{90} ( 7\bs{\mathcal{F}}_1 + 32\bs{\mathcal{F}}_2 + 12\bs{\mathcal{F}}_3 + 32\bs{\mathcal{F}}_4 + 7\bs{\mathcal{F}}_5).
\end{align*}